\documentclass[12pt]{article}
\usepackage{cite}
\newcommand{\Poin}{Poincar{\'e}}
\def\Re{\mathop{\rm Re}\nolimits}
\def\Im{\mathop{\rm Im}\nolimits}

\def\Tr{{\rm Tr}\hskip 1pt}
\def\STr{{\rm STr}\hskip 1pt}
 \newcommand{\ft}[2]{{\textstyle\frac{#1}{#2}}}
 \newcommand{\Ka}{K{\"a}hler}
\newsavebox{\uuunit}
\sbox{\uuunit}
    {\setlength{\unitlength}{0.825em}
     \begin{picture}(0.6,0.7)
        \thinlines
        \put(0,0){\line(1,0){0.5}}
        \put(0.15,0){\line(0,1){0.7}}
        \put(0.35,0){\line(0,1){0.8}}
       \multiput(0.3,0.8)(-0.04,-0.02){12}{\rule{0.5pt}{0.5pt}}
     \end {picture}}
\newcommand {\unity}{\mathord{\!\usebox{\uuunit}}}
\newcommand{\rmi}{{\rm i}}
\newcommand{\rmd}{{\rm d}}
\newcommand{\rme}{{\rm e}}
 \newcommand{\Nphi}{{\cal N} }
\newcommand{\Yrho}{Y}

\csname @addtoreset\endcsname{equation}{section}
\begin{document}
\begin{titlepage}
\begin{flushright}
LPT Orsay--04/15\\
SU-ITP-04/02\\
NYU-TH/04/02/9\\
KUL-TF-04/02\\
hep-th/0402046
\end{flushright}
\vspace{.3cm}
\begin{center}
\baselineskip=16pt
{\LARGE    Fayet--Iliopoulos Terms in Supergravity and Cosmology  
}\\
\vskip 3mm
{\large Pierre Bin{\'e}truy $^1$, Gia Dvali $^2$, Renata Kallosh $^3$ and
Antoine Van Proeyen
$^4$ } \\
\vskip 2mm
{\small $^1$ Laboratoire de Physique Th{\'e}orique, Universit{\'e} Paris-Sud,
F-91405 Orsay, France \\ {\em and} APC, Universit{\'e} Paris 7,  France
\\ \vspace{6pt}
$^2$
Center for Cosmology and Particle Physics, \\
Department of Physics, New York University,\\
New York, NY 10003\\ \vspace{6pt}
$^3$ Department of Physics, Stanford University,\\
Stanford, CA 94305-4060, USA.\\ \vspace{6pt}
$^4$ Instituut voor Theoretische Fysica, Katholieke Universiteit Leuven,\\
       Celestijnenlaan 200D B-3001 Leuven, Belgium.
 }
\end{center}
\vfill
\begin{center}
{\bf Abstract}
\end{center}
{\small We clarify  the structure of  $N=1$  supergravity in 1+3
dimensions with constant Fayet--Iliopoulos (FI) terms. The FI terms
$g\xi$ induce non-vanishing $R$-charges for the  fermions and the
superpotential. Therefore  the $D$-term inflation model in supergravity
with constant FI terms has to be revisited. We present all corrections of
order $g\xi/M_P^2$ to the classical supergravity action required by local
supersymmetry and provide a  gauge-anomaly-free version of the model.

We also investigate the case of the so-called anomalous $U(1)$ when a
chiral superfield is shifted under $U(1)$. In such a case, in the context
of string theory, the FI terms originate from the derivative of the
K{\"a}hler potential and they are inevitably field-dependent. This raises an
issue of stabilization of the relevant field in applications to
cosmology.

The recently suggested equivalence between the $D$-term strings and
D-branes of type $II$ theory shows that brane-anti-brane systems produce
FI terms in the effective $4d$ theory, with the Ramond-Ramond axion
shifting under the $U(1)$ symmetry. This connection gives the possibility
to interpret many unknown properties of D$-\bar{\mbox{D}}$ systems in the
more familiar language of $4d$ supergravity $D$-terms, and vice versa.
For instance, the shift of the axion field in both cases restricts the
possible forms of the moduli-stabilizing superpotential. We provide some
additional consistency checks of the correspondence of $D$-term-strings
to D-branes and show that instabilities of the two are closely related.
Surviving cosmic D-strings of type $II$ theory may be potentially
observed in the form of $D$-term strings of $4d$ supergravity. We study
such string solutions of supergravity with constant FI terms with one
half supersymmetry unbroken and explain some of the puzzling properties
of the zero modes around cosmic strings, such as the difference between
the numbers of fermionic and bosonic modes.

 }\vspace{2mm} \vfill \hrule width 3.cm
{\footnotesize \noindent e-mails: pierre.binetruy@th.u-psud.fr,
gd23@feynman.acf.nyu.edu,\\ kallosh@stanford.edu,
antoine.vanproeyen@fys.kuleuven.ac.be }
\end{titlepage}

\tableofcontents{}

\parskip 5pt
\section{Introduction}

Current cosmological observations suggest that it may be important to
study the effective four-dimensional gravitational theory derivable from
a fundamental theory, like M/string theory. If the effective theory has
local supersymmetry, it is described by $d=4$, $N=1$ supergravity. For
cosmological applications one is interested particularly in any
possibility to find a de Sitter type configuration (dS) with broken
supersymmetry to describe the currently accelerating universe as well as
a slow-roll stage of the early universe inflation.

The potential in $N=1$ supergravity\footnote{We will limit ourselves to
gravitational, vector and chiral multiplets.} is well known: there is an
F-term potential, constructed in a standard way from the K{\"a}hler potential
and superpotential, in some cases there is also a non-trivial $D$-term
potential, derivable in the standard fashion
\cite{Cremmer:1983en,Wess:1992cp,Binetruy:2000zx}. It is extremely
important in the context of cosmological applications that the $D$-term
potential is always positive, whereas the F-term in general has both
positive and negative contributions.

Since the $D$-term potential is positive definite it may lead to de
Sitter type solutions, particularly in presence of a constant FI term.

At present there is no known way to derive the effective  $d=4$, $N=1$
supergravity with constant FI terms from M/string theory. Only
field-dependent $D$-terms have been identified so far (one should keep in
mind that FI terms studied in the context of open string theory are not
immediately relevant for  gravity and cosmology). There is no strict
no-go theorem about the absence of constant FI terms in string theory,
however, for all practical purposes, the situation is close to the
existence of such theorem.\footnote{We are grateful to S. Kachru and J.
Maldacena for numerous discussions of this issue.}

Since constant FI terms in effective supergravity in 1+3 dimensions lead
to dS spaces, the possibility to get such terms from M/string theory may
require new developments in the understanding of string theory. It is
worth reminding here that the second string revolution has allowed to
treat M-theory and 11-dimensional supergravity  as leading to effective
theories with 1+3 dimensional chiral fermions. Before  the
compactifications on  orbifolds and orientifolds were studied, it was
believed that it is impossible to get $d=4$ chiral fermions from
11-dimensional supergravity. At present we may only hope that some new
possibility will realize  in M/string theory that will allow  to derive
constant FI terms and dS and near dS spaces.

The purpose and the results of this paper can be summarized as follows:

\begin{itemize}
\item  Firstly,   we will  study the general case when local $N=1$ supersymmetry in
1+3 dimensions admits constant FI terms and provide the supersymmetry
rules in such theories.\footnote{In globally supersymmetric theories the
constant FI terms can be added   without any  constraints on the theory.
However, in the local case this is not true anymore.} As an application
of these rules we will revisit the $D$-term inflation model and correct
the supergravity version of it to comply with the  restrictions on the
superpotential required when  constant FI terms are present. We will also
study $D$-term strings and their properties in supergravity with constant
FI terms.

\item Secondly, we will study the so-called anomaly generated FI terms originating from string theory
and explain that a procedure of stabilization of certain moduli is
required for these models to be used in the cosmological context in 1+3
dimensions.

\item The recently suggested\cite{Dvali:2003zh} equivalence between the
$4d$ supergravity $D$-term strings and D-branes of type $II$ theory shows
that brane-anti-brane systems in an effective $4d$ theory can be
described as gauge theories with non-zero FI term. The axion shifting
under the $U(1)$-symmetry is dual to the Ramond-Ramond form. This
connection gives the possibility to establish a useful dictionary between
the two descriptions and interpret many important properties of
D$-\bar{\mbox{D}}$ systems in a simpler language of $4d$ supergravity
gauge theories with non-zero $D$-terms, and vice versa. In particular, we
can use our knowledge of the stability of supergravity vacua with
non-zero $D$-terms for understanding the stability of the string vacua
with brane-anti-brane systems and their various cosmological
applications, such as $D$-brane inflation. 
For instance, the shift of the axion in both cases restricts the possible
forms of the moduli-stabilizing superpotential.

 We provide some additional consistency checks of the correspondence between
$D$-term-strings and D-branes and show that, not surprisingly, the
instabilities of the two are closely related. Thus, not only the cosmic
$D$-term-strings, formed after $D$-term inflation, do not cause any
cosmological trouble, but in fact they may be potentially detected.
Hence, the D-brane strings of type $II$ theory, could in principle be
observed in the sky in the form of the supergravity $D$-term strings!
\end{itemize}

A standard expectation is that any K{\"a}hler potential and any holomorphic
superpotential may define a version of $N=1$ supergravity in 1+3
dimensions. We will clarify here  the situation with constant FI terms,
when this expectation is not valid and certain restrictions on the choice
of the superpotential are required.

The very first version of supergravity with  locally supersymmetric
extension of the FI term of the Abelian vector multiplet was constructed
in \cite{Freedman:1976uk}. It has positive cosmological constant,
$\Lambda
>0$.  It was also shown there that local supersymmetry requires the axial
gauging of  gravitino and gaugino (local $R$-symmetry). This theory
involves only the gravitational supermultiplet and the vector
supermultiplet and  has one-loop  axial anomalies  \cite{DG}. More
general classes of models with constant FI terms and scalar
supermultiplets were constructed in
\cite{Stelle:1978wj,Barbieri:1982ac,Ferrara:1983dh,Kallosh:2000ve}. More
recently there were few important developments in studies of some
anomaly-free models with gauged $R$-symmetry and constant  FI terms in
supergravity \cite{Chamseddine:1995rs,Castano:1996ci}. At that time the
main focus of such investigations was towards particle physics with
vanishing cosmological constant.

On the other hand, in the cosmology community, the role of $D$-terms has
become extremely important as a possible origin of de Sitter
configurations and inflation in supergravity
\cite{Binetruy:1996xj,Halyo:1996pp}. It remains, however,  not well known
that the presence of constant FI terms poses specific restrictions on
supergravity theories (see however
\cite{Barbieri:2002ic,Arkani-Hamed:2003mz}).

The existing versions of supergravities with FI terms are mostly
incomplete for our purposes. The $D$-term inflation model has an
important property that in the unstable de Sitter vacuum as well as in
the absolute Minkowski vacuum the superpotential vanishes, $W_{\rm
min}=0$. However, outside the minimum, the superpotential does not
vanish, $W\neq 0$. Thus formulations of supergravity
\cite{Cremmer:1983en,Ferrara:1983dh}, where the Lagrangian depends not on
two functions, the \Ka\ potential ${\cal K}(z, z^*)$ and the
superpotential $W(z)$, but only on one combination ${\cal G}(z, z^*) =
-{\cal K} (z, z^*) - \ln |M_P^{-3}W|^2$ are not suitable\footnote{In
\cite{Ferrara:1983dh} there is a short ``Note added'' how to treat the
case with vanishing superpotential.} since they are not well defined at
$W=0$.

The superspace approach with a non-singular dependence on the
superpotential $W$ presented in \cite{Barbieri:1982ac,Wess:1992cp} has
all terms depending on constant FI. However, the holomorphic kinetic
function $f_{\alpha\beta}(z)$ for the vector multiplets is the simplest
one, equal to 1. On the other hand, in \cite{Binetruy:2000zx} where there
is an arbitrary scalar dependent $f_{\alpha\beta}(z)$, the constant FI
terms are not introduced. The significance of a generic, scalar dependent
$f_{\alpha\beta}(z)$ has to do with axial coupling $aFF^*$ which
sometimes plays an important role in the mechanism of anomaly
cancellation. \bigskip

In section~\ref{ss:superconfAction} we give a summary of the ingredients
of the construction of the supergravity action with superconformal
symmetry. For our purpose it is most useful to study the formulation of
supergravity with the superconformal origin which was recently
constructed in \cite{Kallosh:2000ve}.  It has all 3 generic functions,
the \Ka\ potential ${\cal K}(z, z^*)$, the superpotential $W(z)$, and the
kinetic function $f_{\alpha\beta}(z)$ for the vector multiplets  and the
theory is regular at $W=0$. One furthermore has to define the symmetry
transformations. This includes for any $U(1)$ factor the possible
occurrence of a FI constant $\xi _{\alpha i}$.

In the superconformal approach, one constructs in a first step the action
with full superconformal symmetry. It contains an extra chiral multiplet,
which was often called `compensating multiplet', but was baptized
`conformon' in~\cite{Kallosh:2000ve} to reflect its significance. In the
next step, the gauge symmetries that are not present in Poincar{\'e}
supergravity, such as local dilations, local chiral
$U(1)$-symmetry\footnote{In the context of superconformal $SU(2,2|1)$
symmetry  one often calls the superconformal chiral $U(1)$ symmetry
``$R$-symmetry'', since it rotates the supercharges (transforms the
gravitinos). However, in the Poincar{\'e} supergravity, after the
superconformal $U(1)$ symmetry is fixed, there are sometimes other $U(1)$
symmetries which are  combinations of the superconformal $U(1)$ symmetry
and some additional $U(1)$ gauge symmetries which were present in the
superconformal theory. These  $U(1)$ symmetries  one also calls
$R$-symmetries. In what follows, we will use the term ``local
$R$-symmetries'' in the context of Poincar{\'e} supergravity only.} and local
$S$-supersymmetry, are gauge fixed. This is discussed in the beginning of
section~\ref{s:superconf}.

 The formulation of the theory in \cite{Kallosh:2000ve} has several
advantages. For example, it simultaneously incorporates two different
formulations of phenomenological supergravity depending on the gauge
fixing of the local chiral $U(1)$-symmetry. The first  formulation, in a
\Ka\-covariant gauge, which is more standard, corresponds to
\cite{Cremmer:1983en}. The other one, in a new gauge where the conformon
is real, is closer to \cite{Wess:1992cp,Binetruy:2000zx}, and has a
non-singular dependence on the superpotential $W$.

The new formulation \cite{Kallosh:2000ve}  allowed  to give a detailed
explanation of the superconformal origin of FI-terms by including gauge
transformations of the conformon field as first suggested in
\cite{Ferrara:1983dh}. The conformon field $Y$ is one of the extra
superfields of the superconformal version of the theory, which gets fixed
to remove the local dilatation and local chiral $U(1)$-symmetry. However,
at the superconformal level before the gauge fixing  such a field may
participate in gauge transformations, which turn out to provide the FI
terms:
\begin{equation}
  \delta _\alpha \Yrho =\rmi  {g\xi_\alpha\over 3M_P^2} \Yrho\,\,,\qquad
  \delta _\alpha z_i= \eta_{\alpha i}(z)\,,
\label{delarhoz}
\end{equation}
where $g$ is the gauge coupling constant. When $\xi_\alpha$ are some real
constant terms in some of the $U(1)$'s, they turned out to be constant FI
terms $\xi_\alpha$ in the related $U(1)$ in the supergravity theory.  All
corrections to the supergravity action in such a case can be deduced from
the original superconformal action.

The scaling of fields that allows a suitable rigid limit is discussed in
section~\ref{ss:rigidlimit}. It allows us to present the action and
transformation rules of supergravity with chiral and vector multiplets in
a simple form in section~\ref{ss:simplAction}. The different
contributions to $R$-symmetry and $D$-terms and the implications for the
superpotential are collected in section~\ref{ss:summRFIW}. There the
difference between field-dependent and constant FI terms is clearly
exhibited, but also it is shown how terms can be reinterpreted by
performing K{\"a}hler transformations. The final part of
section~\ref{s:superconf} shows how effective constant FI terms may
result from field-dependent FI terms by replacing a chiral multiplet by
its constant value without breaking local supersymmetry. This procedure
is obvious in rigid supersymmetry, but cannot be done in supergravity in
general. We treat in section~\ref{ss:susyRemoveCh} a case with a K{\"a}hler
potential that splits in two parts.

Section~\ref{s:DtermcstFI} shows how the $D$-term inflation is modified
by this connection to $R$-symmetry. We present here  corrections to the
action  proportional to $g\xi/ M_P^2$, which are required for consistency
of the most  general $N=1$ supergravity with FI terms. We give an
explicit example of such corrections for the supergravity theory
describing $D$-term inflation \cite{Binetruy:1996xj,Halyo:1996pp}. We
also show that such corrections vanish in the limit of rigid
supersymmetry. Such corrections to the $D$-term inflation model have not
been exposed so far. Therefore, we will revisit this model and present a
corrected form of it as an example of the general supergravity with
constant FI terms.

The supergravity theory with constant FI terms has recently been shown to
have $D$-term string solutions with unbroken supersymmetry
in~\cite{Dvali:2003zh}. A short summary of the $D$-term string solution
is presented in section~\ref{ss:Dstringconfig}. In
section~\ref{s:zeromodesD} we study these solutions and their zero modes
in an extended model in which the $D$-term string is coupled to an
arbitrary number of chiral superfields. We present the fermionic zero
modes coming from these superfields and verify explicitly, as well as
deduce from the superalgebra that the bosonic ones are absent, despite
the unbroken supersymmetry.

Then, we turn to field-dependent FI terms in
section~\ref{ss:FIfromanomU1}. We revisit the issues of the $D$-term
inflation model for the case of string theory inspired anomalous FI
terms. We discuss the the relation with the anomaly cancellation by the
Green-Schwarz mechanism.

We show that the true derivation of such cosmological models from string
theory requires to find a stabilization mechanism for the dilaton and/or
volume moduli. We describe some preliminary efforts in this direction
existing in the literature, which may eventually lead to a stringy
version of $D$-term inflation. Furthermore we discuss  the scales of $F$
and $D$-terms.

The cosmological implications of $D$-term strings for D-brane systems and
D-brane inflation are discussed in section~\ref{s:cosmoApplicD}.
Especially the stability is discussed first in the supergravity
formulation (section~\ref{ss:DstabilSG}). Then the relation of the
instabilities of type $II$ D-strings and the $D$-term strings in
supergravity is exhibited in section~\ref{ss:DDinstab}.

The connection of the supergravity $D$-term description to the
D-brane-anti-brane configuration is further deepened in
section~\ref{ss:FIDfromD} by mapping the the moduli stabilization issues
in the two cases.

Appendix~\ref{app:decompsusy} on the residual supersymmetry algebra of
the $D$-term string configuration and appendix~\ref{app:BosFerModes} that
repeats the the relation between fermionic and bosonic modes are useful
for section~\ref{s:zeromodesD}.

\section{Short overview of the local superconformal action}
\label{ss:superconfAction}

We start in \cite{Kallosh:2000ve} with  the $SU(2,2|1)$-invariant action
for $n+1$ chiral multiplets $X_I$  and $(X_I)^*\equiv X^I$ and some
number of Yang--Mills vector multiplets $\lambda^\alpha$ superconformally
coupled to supergravity. The supergravity is represented by a Weyl
multiplet consisting of a vielbein, a gravitino and a gauge field for
$U(1)$ gauge symmetry, gauged $R$-symmetry. The action consists of three
parts, each of them being separately conformally invariant.
\begin{equation}
{\cal L} =  [\Nphi (X,X^*)]_D + [\mathcal{W}(X)]_F + \left[ f_{\alpha
\beta } (X) \bar \lambda_L ^\alpha \lambda_L ^\beta\right] _F
\label{symbL}
\end{equation}

The superconformal action has a number of extra gauge symmetries, by
comparison with ordinary supergravity. The function $\Nphi (X, X^*)$ is
an homogeneous function of degree one in $X$ and $X^*$. Upon gauge-fixing
of extra gauge symmetries it will be related to the \Ka\ potential. The
holomorphic function ${\cal W}(X)$ encodes the superpotential. The
holomorphic function $f_{\alpha \beta}(X)$ encodes the kinetic terms for
the vector multiplet fields.

The extra symmetries include  local dilatation, local $R$-symmetry and
$S$-supersymmetry. In particular, fixing local dilatation removes one of
the chiral scalars so that in supergravity there are only $n$ of them.
Fixing special supersymmetry removes an extra fermion field.  One first
performs a change of variables \cite{Kallosh:2000ve} of the $n+1$
variables $X_I$ to $\Yrho $, which will be the conformon scalar, and $n$
physical scalars $z_i$, which are hermitian coordinates for parametrizing
the \Ka\ manifold in the
 \Poin\ theory. One defines
\begin{equation}
  X_I=\Yrho\, x_I(z_i)\, .
\label{Xrhoxz}
\end{equation}
Here $z_i$, $(z_i)^*\equiv z^i$  are $n$ chiral superfields of ordinary
supergravity and $Y$ is the so-called conformon superfield. The  \Ka\
potential and metric in supergravity are related to the original
superconformal structures as follows:
\begin{eqnarray}
  {\cal K}(z,z^*)&=&-3 \ln \left[-\ft13 \Nphi/(YY^*)\right]\,,
  \nonumber\\
g^i{}_j\equiv  \partial ^i\partial _j {\cal K}&=& -3 (\partial
^iX_I)(\partial _jX^J)\partial
  ^I\partial _J\ln \Nphi\,.
\label{definK}
\end{eqnarray}
In addition to local $SU(2,2|1)$ symmetry, the action may have some
Yang-Mills gauge symmetries when there are Killing symmetries on the
scalar manifold:
\begin{eqnarray}
\delta _\alpha  X_I=k_{\alpha I}(X)&\,, \qquad&
  \delta _\alpha  X^I=k_\alpha{}^I(X^*)\, .
\label{Killingv}
\end{eqnarray}
Here the Killing vectors $k_{\alpha I}(X)$ are holomorphic functions and
their commutators define the structure constants. There are also
corresponding transformations for the vector fields $W_\mu^\alpha$ and
fermionic fields. These Yang-Mills gauge symmetries commute with local
superconformal symmetries. The functions $\Nphi$, ${\cal W}$ should be
invariant and $f_{\alpha \beta }$   covariant under the  Yang-Mills gauge
symmetries, e.g.
\begin{equation}
  \Nphi^Ik_{\alpha I}+\Nphi_Ik_\alpha {}^I={\cal W}^Ik_{\alpha I}=0\,.
\label{invfunct}
\end{equation}
We use here and below the notation where ${\cal N}_I\equiv \partial {\cal
N} /\partial X^I$, and similar for ${\cal N}^I$ and ${\cal W}^I$.

These symmetries of the superconformal action become a main focus of our
attention in the present study of FI terms in supergravity. One finds
that\footnote{We changed  $\xi_{\alpha i}$ in \cite{Kallosh:2000ve} to
$\eta_{\alpha_i}$.}
\begin{equation}
  k_{\alpha I}=\Yrho \left[r_\alpha (z) x_I(z)+\eta_{\alpha i}(z)\partial
  ^ix_I(z)\right] \,,
\label{k_in_r_and_xi}
\end{equation}
where the {\it Yang-Mills transformations of  all chiral superfields in the superconformal action} are
\begin{equation}
  \delta _\alpha \Yrho =\Yrho\, r_\alpha (z)\,,\qquad
  \delta _\alpha z_i= \eta_{\alpha i}(z)\,,
\label{delarhoz2}
\end{equation}
where $r_\alpha (z)$ and $\eta_{\alpha i}(z) $ are $n+1$ holomorphic
functions for every symmetry.

The Yang--Mills gauge transformations of the scalars, which may also act
on the conformon multiplet when $r_\alpha (z)\neq 0$, are the Killing
isometries that are gauged.

The invariance of $\Nphi$ as written in (\ref{invfunct}) leads to
\begin{equation}
  0=\Nphi^Ik_{\alpha I}+\Nphi_Ik_\alpha{}^ I=\Nphi\left[ r_\alpha (z)
  +r^*_\alpha (z^*)-\ft{1}{3}\left(\eta_{\alpha i}
  \partial ^i {\cal K}(z,z^*) +\eta_\alpha {}^i\partial _i{\cal K}(z,z^*)\right)
  \right]\,.
\label{preKillingEqn}
\end{equation}
Thus $r_\alpha(z)$ describes the non-invariance of the \Ka\
potential:
\begin{equation}
  \delta _\alpha {\cal K}=\eta_{\alpha i}(z)\partial ^i{\cal K}(z,z^*)
  +\eta _\alpha {}^i\partial _i{\cal K}(z,z^*)=
  3(r_\alpha(z)+r_\alpha^*(z^*))\,.
\label{deltaalphaK}
\end{equation}
Imaginary constants in $r_\alpha $ do not show up here. In the special
case that the transformation of the conformon superfield $Y$ is given by
imaginary constants (this is consistent with the YM algebra for $U(1)$
factors)
\begin{equation}
r_\alpha = \rmi{g\xi_\alpha\over 3 M_P^2}\, , \qquad  \partial _i
\xi_\alpha =0\,,
\end{equation}
one finds that the \Ka\ potential is invariant,
\begin{equation}
  \delta _\alpha {\cal K}=0\,.
\label{InvK}
\end{equation}
The vector multiplets have the auxiliary field $D^\alpha $, whose value
is given by
\begin{eqnarray}
 D^\alpha &=&\left(\Re f_{\alpha \beta }\right) ^{-1} {\cal P}_\beta +
\mbox{ fermionic terms ,}\nonumber\\
{\cal P}_\alpha(z,z^*) &=&\ft12\rmi M_P^2\left[\left(
 \eta_{\alpha i}(z)\partial ^i{\cal K}(z,z^*)-
\eta_\alpha {}^i\partial _i{\cal K}(z,z^*)\right) -3 r_\alpha(z)+3
r_\alpha^*(z^*))\right]\nonumber\\ &=& \rmi\,M_P^2\left( -\eta_\alpha
{}^i\partial _i{\cal K}(z,z^*)+3r_\alpha^*(z^*)\right)=\rmi\,M_P^2\left(
\eta_{\alpha i}\partial ^i{\cal K}(z,z^*)-3r_\alpha(z)\right).
\label{Pgeneral}
\end{eqnarray}
These real functions $ {\cal P}_\alpha(z,z^*)$,  called Killing
potentials, encode the Yang-Mills transformations. Indeed, their derivatives
determine the Killing vectors:
\begin{equation}
\partial _i{\cal P}_\alpha(z,z^*)=\rmi\,M_P^2\,\eta_{\alpha j}g^j{}_i \,.
\label{Killvecfrpot}
\end{equation}
In the case where $r_\alpha = \rmi g\xi_\alpha / (3 M_P^2)$,
\begin{eqnarray}
 {\cal P}_\alpha(z,z^*) = -\rmi\,M_P^2 \eta_\alpha
{}^i\partial _i{\cal K}(z,z^*)+g \xi_\alpha = \rmi\,M_P^2 \eta_{\alpha i}
\partial ^i{\cal K}(z,z^*)+g \xi_\alpha \,. \label{P}
\end{eqnarray}
The properties of ${\cal W}$ under chiral and dilatational symmetries imply that it is of
the form
\begin{equation}
   {\cal W}=\Yrho ^3 M_P^{-3}W(z)\,.
\label{superpotential}
\end{equation}
Invariance of ${\cal W}$ under YM transformations,
\begin{equation}
  \delta _\alpha {\cal W}=0\,,
\label{InvcalW}
\end{equation}
requires that
\begin{equation}
  \eta _{\alpha i}\partial ^iW=-3r_\alpha W\,.
\label{gtransfW}
\end{equation}
Thus, one finds that the superpotential $W$ cannot be Yang-Mills
invariant (which would correspond to $ \eta_{\alpha i}\partial ^iW=0$) in
all models where the conformon multiplet transforms under gauge
transformations. In particular, when $r_\alpha=\rmi
g\xi_\alpha/(3M_P^2)$, which will be shown later to correspond to a
constant FI term in supergravity $\xi_\alpha$, we find
\begin{equation}
 \eta_{\alpha i}\partial ^iW=- \rmi {g\xi_\alpha\over M_P^2} W\,.
\label{FIgtransfW}
\end{equation}

\section{From superconformal theory to supergravity with FI }
\label{s:superconf}

The gauge fixing of the local dilatational invariance introduces the mass
scale $M_P$ which was absent in the superconformal theory.\footnote{We
use $M_P \equiv M_\mathit{Planck}/\sqrt{8\pi} \sim 2 \times
10^{18}$~GeV.} It also fixes $|Y|$ in terms of a \Ka\ potential depending
only on the physical scalars $z,\, z^*$,
\begin{equation}
  \Yrho \Yrho ^*  \exp \left(- \ft{1}{3}{\cal K}(z, z^*)\right) =M_P^2 = - \ft13{\cal N}
  \,.
\label{rho}
\end{equation}
The \Ka\ invariance has its origin in the non-uniqueness of the
splitting (\ref{Xrhoxz}). This creates an invariance under a
redefinition
\begin{equation}
  \Yrho'= \Yrho\, e^{\ft13\Lambda _\Yrho (z)}\,,\qquad  x'_I=x_I\,
   e^{-\ft13\Lambda _\Yrho   (z)}
\label{firstKatr}
\end{equation}
for an arbitrary holomorphic function $\Lambda _\Yrho (z)$. This
redefinition changes the \Ka\ potential to
\begin{equation}
{\cal K}'={\cal K}+\Lambda_\Yrho  (z)+\Lambda _\Yrho ^*(z^*)\,.
\label{KatrK}
\end{equation}
The \Ka\ transformations act therefore on the superpotential $W$ as
\begin{equation}
  W'=We^{-\Lambda _\Yrho (z)}\,,
\label{KatrW}
\end{equation}
leaving $ {\cal W}$ invariant. The $U(1)$ invariance can be fixed in case
of an always non-vanishing gravitino mass  by choosing
\begin{equation}
 \mbox{\Ka\ symmetric } U(1)-\mbox{gauge }\ :\quad {\cal W}={\cal W}^*\,.
\label{U1gauge}
\end{equation}
This choice of gauge fixing of the chiral $U(1)$   leads to the action of
phenomenological $N=1$ supergravity as given in \cite{Cremmer:1983en}.
This gauge makes sense  only for ${\cal W}\neq 0$, as for ${\cal W}= 0$
the condition is empty. An alternative gauge was suggested in
\cite{Kallosh:2000ve} for the theories where ${\cal W}=0$ at some points
in field space
\begin{equation}
 \mbox{ non-singular at  ${\cal W}=0$ } U(1)-\mbox{gauge }\ :\quad \Yrho =\Yrho^*\,.
\label{U1gauge2}
\end{equation}
In this gauge  for the $U(1)$ symmetry, the theory
 is  non-invariant under the \Ka\ transformations
(\ref{firstKatr}). This implies that the remaining invariance is a
combination of chiral $U(1)$ and \Ka\ transformations. The action in this
form will be closer to the action of the phenomenological N=1
supergravity as given in \cite{Wess:1992cp,Binetruy:2000zx}, where it was
derived by superspace methods.

We will now give the action, including the fermions. First we fix the
notation. The gravitinos are usually written as Majorana spinors $\psi
_\mu $, but sometimes it is convenient to split them into their complex
chiral parts,
\begin{equation}
  \psi _{\mu L}=\ft12(1+\gamma _5)\psi _\mu \,,\qquad\psi _{\mu R}=\ft12(1-\gamma _5)\psi _\mu
  \,.
 \label{psiLR}
\end{equation}
The same notation applies to the gauginos $\lambda ^\alpha $. The spinors
of the chiral multiplets are always denoted by their chiral parts. We use
the position of the index $i$ to indicate the chirality, with
\begin{equation}
\chi_i =\ft12(1+\gamma_5) \chi_i\,, \qquad \chi^i = \ft12(1-\gamma_5)
\chi^i\,.
 \label{chiLR}
\end{equation}

The action can be written as
\begin{eqnarray}
e^{-1}{\cal L}&=&-\ft12M_P^2\left[ R + \bar \psi_\mu \gamma ^{\mu \rho \sigma }{\cal
D}_\rho \psi _\sigma \right]
-g_i{}^j\left[M_P^2(\hat{\partial }_\mu z^i)(\hat{\partial }^\mu
z_j)+
 \bar \chi _j  \not\!\! {\cal D} \chi^i+ \bar \chi^i
 \not\!\! {\cal D} \chi_j \right]\nonumber\\
&+&(\Re f_{\alpha \beta})\left[ -\ft14 F_{\mu \nu }^\alpha F^{\mu \nu
\,\beta } -\ft12 \bar \lambda ^\alpha \not\!\!\hat{\cal D}\lambda
^\beta \right] +\ft 14\rmi(\Im f_{\alpha \beta})\left[
 F_{\mu \nu }^\alpha \tilde F^{\mu \nu \,\beta }- \hat{\partial }_\mu\left(
\bar \lambda ^\alpha \gamma _5\gamma ^\mu \lambda
^\beta\right)\right] \nonumber\\
 &-&M_P^{-2} e^{\cal K}\left[ -3 WW^*+({\cal D}^iW) g^{-1}
 {}_i{}^j({\cal D}_jW^*)\right] -\ft{1}{2}(\Re f)^{-1\,\alpha \beta }
 {\cal P}_\alpha {\cal P}_\beta
 \nonumber\\
&+&\ft18(\Re f_{\alpha \beta})\bar \psi _\mu \gamma ^{\nu \rho
}\left( F_{\nu \rho }^\alpha+ \hat F_{\nu \rho }^\alpha \right)
\gamma ^\mu \lambda ^\beta \nonumber\\
&+& \left\{M_Pg _j{}^i \bar
\psi_{\mu L}(\hat{\not\! \partial } z^j)
 \gamma^\mu \chi_i
 + \bar  \psi_R  \cdot \gamma\left[\ft12\rmi\lambda_L^\alpha
 {\cal P}_\alpha
+\chi _i  e^{{\cal K}/2}M_P^{-1} {\cal D}^iW  \right]\right.
  \nonumber\\
 &&\ + \ft12  e^{{\cal K}/2}W\bar \psi _{\mu R} \gamma ^{\mu \nu }\psi _{\nu R}
- \ft14M_P^{-1}f^i_{\alpha \beta}\bar \chi _i\gamma ^{\mu \nu } \hat F_{\mu
\nu }^{-\alpha } \lambda _L^\beta \nonumber\\
 && \
 -e^{{\cal K}/2} M_P^{-2}({\cal D}^i{\cal D}^j  W)\bar \chi _i\chi _j
 +\ft 12\rmi(\Re f)^{-1\,\alpha \beta} {\cal P}_\alpha  M_P^{-1}f^i_{\beta \gamma }
\bar \chi _i\lambda ^\gamma
  - 2 M_P g \xi _\alpha{}^i g_i{}^j  \bar \lambda ^\alpha \chi_j
 \nonumber\\
  &&\ + \ft14 M_P^{-2} e^{{\cal K}/2}({\cal D}^j W) (g^{-1})_j{}^i
  f_{\alpha \beta i}\bar \lambda_R ^\alpha \lambda_R ^\beta
 + {\rm h.c.} \} +\mbox{4-fermion terms} \,,
\label{phenomL}
\end{eqnarray}
where
\begin{eqnarray}
    \hat{F}^\alpha_{\mu
\nu }&=&F^\alpha_{\mu \nu }+\bar \psi _{[\mu }\gamma _{\nu ]}\lambda
^\alpha\,,\qquad F^\alpha_{\mu \nu }=2\partial _{[\mu }W^\alpha _{\nu
]}+W_\mu ^\beta W_\nu ^\gamma f_{\beta \gamma }^\alpha\,, \nonumber\\
\tilde F^{\mu \nu\,\alpha  }&=& \ft12 e^{-1}\varepsilon ^{\mu \nu \rho
\sigma } F^\alpha_{\rho \sigma }\,,\nonumber\\ \hat{F}^{-\alpha }_{\mu
\nu }&=&\ft12\left(\hat{F}^\alpha_{\mu \nu }
-\tilde{\hat{F}}{}^\alpha_{\mu \nu }\right) = F^{-\alpha }_{\mu \nu
}-\ft14 \bar \psi _{\rho L}\gamma _{\mu \nu }\gamma ^\rho \lambda ^\alpha
_R +\ft14 \bar \psi_R\cdot \gamma\gamma _{\mu \nu }\lambda ^\alpha _R \,, \nonumber\\
{\cal D}^iW&=& \partial  ^i W+\left( \partial ^i {\cal K}\right) W \,, \nonumber\\
 {\cal D}^i{\cal D}^jW&=&\partial ^i{\cal D}^jW+
  \left( \partial ^i {\cal K}\right) {\cal D}^jW-\Gamma ^{ij}_k{\cal D}^k W\,,
\qquad
   \Gamma_k^{ij} = (g^{-1})_k{}^\ell \partial ^ig^j_\ell\,.
   \label{covcalD}
\end{eqnarray}

We have skipped  the 4-fermion terms here, referring the reader to eq. (5.15) in \cite{Kallosh:2000ve}. These terms will not be affected by the presence of constant FI terms. Here
$g_i{}^j$ is the \Ka\ metric, see (\ref{definK}).
The covariant derivative of $z$ is
\begin{equation}
  \hat{\partial }_\mu z_i= \partial _\mu z_i -W_\mu ^ \alpha \eta_{\alpha i}\,.
\label{hatdz}
\end{equation}
The covariant derivatives on gaugino and gravitino are
\begin{eqnarray}
{\cal D}_\mu \lambda ^\alpha &=&\left( \partial _\mu +\ft14 \omega
_\mu {}^{ab}(e)\gamma _{ab} +\ft 12\rmi A^B_\mu \gamma _5\right)
\lambda ^\alpha  -W_\mu ^\gamma \lambda ^\beta f_{\beta \gamma
}^\alpha \,,    \nonumber\\  {\cal D}_{[\mu }\psi _{\nu ]}&=&\left(
\partial _{[\mu} +\ft14 \omega _{[\mu} {}^{ab}(e)\gamma _{ab} +\ft
12\rmi A^B_{[\mu} \gamma _5\right)\psi _{\nu ]}\,. \label{covderAB}
\end{eqnarray}
where the $U(1)$ connection $A_\mu ^B$ in our gauge $Y=Y^*$ is given by
\begin{equation}
  A_\mu ^B=\frac{1}{2}\rmi\left[ (\partial _i{\cal K})\partial _\mu z^i-(\partial
  ^i{\cal K})\partial _\mu z_i\right]
  +\frac{1}{M_P^2}W_\mu ^\alpha {\cal P}_\alpha \,.
  \label{AmuBinz}
\end{equation}
In the superconformal theory, this field was the gauge field of the
$U(1)$ $R$-symmetry of the superconformal algebra, but here it is simply
an auxiliary field. The covariant derivative on the fermions of the
chiral multiplets $\chi _i$ is
\begin{equation}
  {\cal D}_\mu \chi _i=\left( \partial _\mu +\ft14\omega _\mu
  {}^{ab}(e)\gamma _{ab}-\ft 12\rmi A_\mu ^B \gamma_5\right) \chi _i+\Gamma _i^{jk}\chi
  _j \hat{\partial }_\mu z_k  -W_\mu ^\alpha \left( \partial
  ^j\eta _{\alpha i}\right) \chi _j\,,
\label{Dchii}
\end{equation}
where the \Ka\ connection (\ref{covcalD}) has been used.

The action (\ref{phenomL}) is invariant under the local \Poin\ group and
$Q$-supersymmetry, which are standard local symmetries of supergravity.
In addition, there are also some gauge symmetries with gauge fields
$W_\mu ^\alpha$. We concentrate further on $U(1)$ factors. Under these
symmetries the gravitino, gaugino and chiral fermions are charged when
the constant FI terms $\xi_\alpha$ are present in ${\cal P}_\alpha$.

Thus there are quite a few places where FI terms appear. They obviously
appear in the potential and in covariant derivatives.  Less obviously,
they also appear via $\eta_{i\alpha}$ since a particular combination of
Killing vectors must satisfy the condition (\ref{FIgtransfW}) that the
superpotential $W$ transforms under $U(1)$'s.


In the usual context of K{\"a}hler geometry, Killing potentials are
determined up to constants (see e.g.\ \cite{Wess:1992cp}, appendix~D),
since only the differential equation of the type shown in
(\ref{Killvecfrpot}) is available. The superconformal approach and that
of~\cite{Wess:1992cp} have a different starting point.
In  \cite{Wess:1992cp} the arbitrary constants in the solution of the
differential equation (\ref{Killvecfrpot}) are the cause of the FI terms.
In \cite{Kallosh:2000ve}, as first recognized in \cite{Ferrara:1983dh},
the gauge transformations of the conformon multiplet, encoded in
$r_\alpha (z)$, are responsible for the FI terms. These gauge
transformations then induce constants in ${\cal P}_\alpha $.

First observe that $r_\alpha \neq 0$ signals the mixture of chiral
transformations and  gauge transformations with  index $\alpha $. Indeed,
after fixing the modulus of $\Yrho $ by some gauge choice, the remaining
invariance is the linear combination of gauge transformations that leaves
$\Yrho $ invariant, and this depends on $r_\alpha $. Another way to see
this is that the gravitino field couples to $A_\mu ^B$.

A short summary of the $R$-symmetry charges in supergravity with constant
FI terms is the following. For gravitino and gaugino we have the axial
coupling with some vector fields $W_\mu^\alpha$, and the couplings are
proportional to $ G_{\alpha}\equiv\xi_\alpha/(2 M_P^2) $:
\begin{eqnarray}
{\cal D}_\mu \lambda ^\alpha &=&\left( \partial _\mu  +\rmi g G_{\beta}
W^\beta_\mu \gamma _5 +\cdots \right) \lambda ^\alpha   \,, \nonumber\\
{\cal D}_{[\mu }\psi _{\nu ]}&=&\left(
\partial _{[\mu}  +\rmi gG_{\beta} W^\beta_{[\mu} \gamma _5+\cdots \right)
\psi _{\nu ]}\,, \label{Rcharges}\\
{\cal D}_\mu \chi _i&=&\left( \partial _\mu -\rmi g G_{\beta} W^\beta_\mu
 +\cdots \right) \chi _i \nonumber \\
&&- W_\mu ^ \alpha \eta_{\alpha k}\Gamma _i^{jk}\chi  _j  -W_\mu ^\alpha
\left( \partial  ^j\eta _{\alpha i}\right) \chi _j\,, \label{Rchiralf}
\end{eqnarray}
For the scalars in the chiral multiplet and for the superpotential we
have
\begin{eqnarray}
  \hat{\partial }_\mu z_i&=& \partial _\mu z_i -W_\mu ^ \alpha \eta_{\alpha i}\,,\nonumber\\
\eta _{\alpha i}\partial ^iW&=&- 2 \rmi g G_{\alpha} W\,.
\label{Rchirals}
\end{eqnarray}

\subsection{Supergravity and a rigid limit}\label{ss:rigidlimit}
The limit from a supergravity theory to a supersymmetry theory is not
always obvious. In many versions of supergravity a notation $M_P=1$ was used which
does not make such limit easy.
The purpose of this subsection is to show how
 to parametrize the scalar fields in a way that the rigid limit can be
taken easily.
 For $N=1$, the following procedure has been proposed in  \cite{Kallosh:2000ve} for the  simplest situation  in which all
scalar fields appear in the rigid limit.

A convenient choice  is to  expand around the point $z=z^0$ and take the special coordinates
\begin{equation}
z_i= z_i^0+ M_P^{-1}\phi_i\,,
\label{zepsphi}
\end{equation}
 with in~(\ref{Xrhoxz})
\begin{equation}
  x_0=1\,,\qquad x_i=M_P^{-1}\phi _i\,
\label{specCoor}
\end{equation}
and
the \Ka\ potential in a form
\begin{equation}
  {\cal K}=
  M_P^{-2}  K(\phi ,\phi ^*,M_P^{-1} )\,,
\label{expK}
\end{equation}
where
$K(\phi ,\phi ^*,M_P^{-1})$ is regular at $M_P^{-1}=0 $.
Note that the \Ka\ metric is
\begin{equation}
  g^i{}_j =\frac{\partial }{\partial z_i}\frac{\partial }{\partial z^j}{\cal
  K}=\frac{\partial }{\partial \phi _i}\frac{\partial }{\partial \phi ^j} K\,,
\label{gijisgij}
\end{equation}
i.e., it does not change under the reparametrizations.
Therefore the
kinetic term for the scalars in (\ref{phenomL}) will loose its
dependence on $M_P$ by the reparametrization as chosen in
(\ref{zepsphi}) which was  the motivation for the
proportionality factor $M_P^{-1}$ in (\ref{zepsphi}). It follows that
 the fields $\phi _i$ have again the same
dimension as the conformal fields $X_I$. One finds that $\Yrho = M_P \exp
[K/(6M_P^{2})]$ and
\begin{equation} X_0=\Yrho =M_P +{\cal O}(M_P^{-1})\,,\qquad
  X_i=\Yrho x_i= \Yrho M_P^{-1}\phi _i= \phi _i\exp [K/(6M_P^{2})] = \phi
  _i +{\cal O}(M_P^{-2})\,.
\label{Xispecial}
\end{equation}
So the fields $\phi _i$ are in lowest order of $M_P^{-1}$ equal to
the conformal fields that we started from. The same happens  with
the conformal fermions.

{}From now on, we will thus use the fields $\phi _i$ and complex
conjugates $\phi ^i$ rather than the $z_i$ and $z^i$ to indicate the
scalar fields. Therefore, \emph{derivatives $\partial ^i$ will  stand for
derivatives w.r.t. $\phi_i$ rather than w.r.t. $z_i$}. The difference is
thus a factor $M_P$. E.g., from now on,
 $f^i_{\alpha \beta }=\partial f_{\alpha \beta}/ \partial \phi _i=
 M_P^{-1}\partial f_{\alpha \beta}/ \partial z_i$.
Thus to use the equations of the previous section we have to replace
\begin{equation}
  \partial ^i\rightarrow M_P\partial ^i\,,\qquad f^i_{\alpha \beta }\rightarrow
  M_Pf^i_{\alpha \beta }\,,\qquad {\cal D}^i\rightarrow M_P{\cal D}^i\,.
 \label{replacements}
\end{equation}
One checks also that
\begin{eqnarray}
\eta _{\alpha i} & = & \delta _\alpha z_i=M_P^{-1}\delta _\alpha \phi _i\,,
 \nonumber\\
\frac{\partial }{\partial z_i}{\cal K} & = & M_P^{-1} \frac{\partial
}{\partial \phi_i} K\,. \label{rigidlimitP}
\end{eqnarray}
One can also check that $ {\cal P}_\alpha $ in (\ref{Pgeneral}) has a
finite rigid limit
\begin{equation}
  {\cal P}_\alpha =\rmi (M_P\eta _{\alpha i})\partial ^iK -3\rmi (M_P^2r_\alpha
  )\,,
 \label{calPnewcoord}
\end{equation}
under the condition that
\begin{equation}
  r_\alpha(z)= M_P^{-2}\tilde r_\alpha(\phi, M_P^{-1})\,,
 \label{tilder}
\end{equation}
where $\tilde r_\alpha(\phi, M_P^{-1})$ is regular at $M_P^{-1}=0 $.

In the limit of rigid symmetries when $M_P\rightarrow \infty$ and the FI
terms $\xi_\alpha$ stay fixed, the axial coupling $G_{\alpha}= \xi_\alpha
/ (2 M_P^2)$ vanishes. The only $\xi_\alpha$-dependent term which
survives this limit is the term in the potential ${1\over 2}g^2
\xi_\alpha^2$. Thus the case of supersymmetric gauge theories, which
allows any number of $U(1)$ groups with FI terms, is obtained in the
proper limit from supergravity.

Consider a \emph{simple example} with (\ref{specCoor}), a trivial K{\"a}hler
metric and a single $U(1)$ group ($\alpha $ takes only one value, and we
write therefore $\xi $ for $\xi _\alpha $), under which the scalars have
a charge $q_i$:
\begin{eqnarray}
  K&=&\phi_i \phi^i\,, \qquad f_{\alpha \beta }=\delta
_{\alpha \beta } \nonumber\\
\delta_\alpha  \phi_i&=& \rmi g q_i \phi_i \,,\qquad \mbox{i.e.}\qquad
\eta _{\alpha i}=\rmi g M_P^{-1}q_i \phi_i\,,\nonumber\\
r_\alpha &=&\ft13\rmi g\xi M_P^{-2}=\ft23\rmi g\,G\,, \qquad G \equiv
\ft12\xi M_P^{-2}\,.
\end{eqnarray}
Then the superpotential ${\cal W}$ has to be gauge-invariant, and with
(\ref{superpotential}),
\begin{equation}
\delta W= \rmi \sum_{i=1}^n g q_i \phi_i \partial ^i W(\phi) =  - \rmi
{g\xi\over M_P^2} W (\phi)= - 2 \rmi g G W (\phi)\,. \label{RinvariantW}
\end{equation}
If this is the case, the action is invariant and we find
\begin{equation}
   {\cal P}= g\left( \xi-\sum_{i=1}^n q_i  \phi_i \phi^i \right) \,.
    \label{Pexample1}
\end{equation}
The potential gets a contribution $\ft12 {\cal P}^2$, which yields the
Fayet--Iliopoulos cosmological constant $\ft12g^2 \xi ^2$.

An important constraint on the combination of charges which appears in a
superpotential of the form $W=\lambda \phi_1 \cdots \phi_n$ is
\begin{equation}
\sum_{i=1}^{i=n} q_i = -{\xi\over M_P^2} = -2G \,.
 \label{sumrule}
\end{equation}
This property is in agreement with potentials in gauge theory being
invariant under gauge transformations since in the limit $M_P \rightarrow
\infty $ the sum of charges vanishes. However, in supergravity the sum of
charges cannot vanish, which will lead to a particular correction in the
supergravity potential for $D$-term inflation.


\subsection{Simplified action}\label{ss:simplAction}

In this section we are  trying to  present the rules for the most general
case of $N=1$ supergravity with constant FI terms and make this section
completely self-contained. For all practical purposes, if one is not
interested in the reasons behind the rules, one should find here all
information for the generic case.

The supergravity action is defined by $W(\phi)$, $K(\phi,  \phi^*)$ and
$f_{\alpha \beta}(\phi)$ as usual. Some of the $U(1)$ gauge groups may
contain constant FI terms $\xi_{\alpha}$. In this section we are
considering a simple case when in the superconformal theory $\tilde
r_\alpha (\phi) = \rmi g\xi_\alpha / 3$. The bosonic part of the action
is
\begin{eqnarray}
e^{-1}{\cal L}_{\rm bos}&=&-\ft12M_P^2 R -g_i{}^j(\hat{\partial }_\mu
\phi ^i)(\hat{\partial }^\mu \phi _j) -V\nonumber\\ [2mm]
 &&-\ft14(\Re f_{\alpha \beta}) F_{\mu \nu }^\alpha F^{\mu \nu \,\beta }
 +\ft 14\rmi(\Im f_{\alpha \beta})
 F_{\mu \nu }^\alpha \tilde F^{\mu \nu \,\beta }\,.
 \label{bosonic}
\end{eqnarray}
The potential consists of an $F$-term and a $D$-term:
\begin{eqnarray}
V&=&V_F+V_D\,,\nonumber\\
V_F &=&\rme^{(K/M_P^2)}\left[ ({\cal D}^iW)(g^{-1})
 {}_i{}^j({\cal D}_jW^*) -3M_P^{-2}WW^*\right]\,,\nonumber\\
  V_D&=&\ft12\left.(\Re f_{\alpha \beta}) D^\alpha D^\beta\right|_{bos}=
 \ft{1}{2}(\Re f)^{-1\,\alpha \beta }  {\cal P}_\alpha {\cal P}_\beta\,,
\label{Vtotal}
\end{eqnarray}
where
\begin{eqnarray}
{\cal D}^iW&=& \partial ^i W +M_P^{-2}(\partial ^i K) W\,, \nonumber\\
 {\cal P}_\alpha(\phi,\phi^*, M_P^{-1})  &=& \rmi\,\left[ M_P\eta_{\alpha i}
\partial^i K(\phi,\phi^*, M_P^{-1})-3 \tilde r_\alpha(\phi,M_P^{-1} )\right]
\nonumber \\
 &=& \rmi\,\left[ -M_P\eta_\alpha{}^i
\partial_i K(\phi,\phi^*, M_P^{-1})+3 \tilde r_\alpha^*(\phi^*,
M_P^{-1} )\right] \label{Pnew} \, ,
\end{eqnarray}
where $M_P\eta_{\alpha i} = \delta_\alpha \phi_i$ and $M_P\eta_\alpha
{}^i = \delta_\alpha \phi^i$.

Under $U(1)$ gauge transformations in  the directions in which there are
FI terms $\xi_\alpha$, the superpotential must transform as
\begin{equation}
\delta_\alpha W= -\rmi {g\xi_\alpha\over M_P^2} W (\phi)\,.
\label{transfW}
\end{equation}

The part of the action quadratic in fermions reads
\begin{eqnarray}
e^{-1}{\cal L}_{\rm fer}&=&- \ft12 M_P^2\bar \psi_\mu \gamma ^{\mu \rho
\sigma }{\cal D}_\rho \psi _\sigma+ \ft12 m\,\bar \psi _{\mu R} \gamma
^{\mu \nu }\psi _{\nu R} +\ft12 m^*\bar \psi _{\mu L} \gamma ^{\mu \nu
}\psi _{\nu L}
\nonumber\\
& -&g_i{}^j \left[\bar \chi _j  \not\!\! {\cal D} \chi^i+ \bar \chi^i
 \not\!\! {\cal D} \chi_j \right]
  - m^{ij}\bar \chi _i\chi _j
 -  m_{ij}\bar \chi ^i\chi ^j
+e^{-1}{\cal L}  _{mix} \nonumber\\
&-& 2 m_{i\alpha} \bar \chi ^i\lambda^\alpha -
2 m^i{}_\alpha \bar \chi _i\lambda^\alpha
- m_{R,\alpha \beta } \bar \lambda_R ^\alpha \lambda_R ^\beta
- m_{L,\alpha \beta } \bar \lambda_L ^\alpha \lambda_L ^\beta
 \nonumber\\
&+&(\Re f_{\alpha \beta})\left[  -\ft12 \bar \lambda ^\alpha \not\!\!{\cal D}\lambda ^\beta
\right] +\ft 14\rmi(\Im f_{\alpha \beta})\left[
 - \hat{\partial }_\mu\left(
\bar \lambda ^\alpha \gamma _5\gamma ^\mu \lambda
^\beta\right)\right] \nonumber\\ &+&\ft14\left\{ (\Re f_{\alpha
\beta})\bar \psi _\mu \gamma ^{\nu \rho } F_{\nu \rho }^\alpha \gamma
^\mu \lambda ^\beta
 - \left[ f^i_{\alpha \beta}\bar \chi _i\gamma ^{\mu \nu }
 F_{\mu \nu }^{-\alpha } \lambda _L^\beta +h.c.\right] \right\},
 \label{shortphenL}
\end{eqnarray}
where
\begin{equation}
  m  =e^{{\cal K}/2}W\,,
\label{defm}
\end{equation}
which is related to the (real) gravitino mass,
$
  m_{3/2}  =|m|M_P^{-2}.
$
Also the following notation are used
\begin{eqnarray}
&&m^i\equiv {\cal D}^i m\phantom{^*}  =   e^{{\cal K}/2} {\cal D}^i W\phantom{^*}=
 \partial ^i m\phantom{^*} +\ft12 (\partial ^i {\cal K}) m\phantom{^*}\,,\qquad
{\cal D}_i m\phantom{^*} \, =  \, \partial _i m\phantom{^*}
-\ft12 (\partial _i {\cal K}) m\phantom{^*} =0\,,\nonumber\\
&&m_i\equiv {\cal D}_i m^*  =   e^{{\cal K}/2} {\cal D}_i W^*=
 \partial_i m^* +\ft12 (\partial _i {\cal K}) m^*\,,\qquad
{\cal D}^i m^* \, =  \, \partial ^i m^*
-\ft12 (\partial^i {\cal K}) m^* =0\, , \nonumber\\
&&\label{defmima}
\end{eqnarray}
and
\begin{eqnarray}
m^{ij} & = &  {\cal D}^i{\cal D}^j  m
= \left(\partial ^i+\ft12 (\partial ^i {\cal K})\right)
m^j-\Gamma ^{ij}_k m^k\,,
\nonumber\\
m_{i\alpha }&=&-\rmi\left[ \partial _i{\cal P}_\alpha -\ft 14 (\Re
f)^{-1\, \beta\gamma} {\cal P}_\beta f_{\gamma\alpha\,i  }\right]\,,
 \nonumber\\
m_{R,\alpha \beta }&=&-\ft14 f_{\alpha \beta i}  (g^{-1})^i{}_j\, m^j \,.
\label{fermionmasses}
\end{eqnarray}
 ${\cal L}_{mix}$ can be written in different ways:
\begin{eqnarray}
e^{-1}{\cal L}  _{mix}&=&
g _j{}^i \bar \psi_{\mu L}(\hat{\not\! \partial } \phi^j)
 \gamma^\mu \chi_i+ \bar  \psi_R  \cdot \gamma\upsilon^1_L+{\rm h.c.},
 \nonumber\\
&=& 2g_j{}^i\bar \psi_{\mu R}\gamma ^{\nu \mu}\chi ^j\hat{\partial }_\nu
\phi_i+\bar  \psi_R  \cdot \gamma\upsilon_L+{\rm h.c.} \,,
\label{Lmixresc}
\end{eqnarray}
where
\begin{eqnarray}
  \upsilon_L&=& \upsilon^1_L+\upsilon^2_L\,,\nonumber\\
\upsilon^1_L&=& \ft12\rmi {\cal P} _\alpha \lambda _L^\alpha +m^i
\chi_i\,,\qquad \upsilon^2_L=
 (\not\!\hat{\partial }\phi_i) \chi ^jg_j{}^i  \,. \label{Goldstino}
\end{eqnarray}
The covariant derivatives on the scalar fields still contain gauge
connection, while the one on the fermions $\chi _i$ contain also Lorentz,
gauge and \Ka\ connections:
\begin{eqnarray}
  {\cal D}_\mu \chi _i&=&\left( \partial _\mu +\ft14\omega _\mu
  {}^{ab}(e)\gamma _{ab}\right) \chi _i -W_\mu ^\alpha \chi _j \partial
  ^j\eta _{\alpha i}   -\frac{\rmi}{2M_P^2}W_\mu ^\alpha {\cal
  P}_\alpha\chi_i
  \nonumber\\ &&+\ft14\left[ (\partial _j{\cal K}){\partial}_\mu \phi ^j -
  (\partial ^j{\cal K}){\partial}_\mu \phi _j\right] \chi _i
  +\Gamma _i^{jk}\chi
  _j \hat{\partial }_\mu \phi_k \,,
\label{Dmuchi}
\end{eqnarray}
where
\begin{equation}
  \partial   ^j\eta_{\alpha i}=\frac{\partial }{\partial z_j}\delta _\alpha
  z_i=\frac{\partial }{\partial \phi_j}\delta _\alpha\phi_i\,.
\label{deljchii}
\end{equation}

The parts of the supersymmetry transformation laws of the fermions
where they transform to bosons, and boson transformations
linear in fermions, are:
\begin{eqnarray}
\delta e_\mu ^a&=&\ft12\bar \epsilon\gamma ^a\psi _\mu \,,\qquad
\delta \phi _i=\bar \epsilon _L\chi _i\,,\qquad
\delta W_\mu ^\alpha =-\ft12\bar \epsilon \gamma _\mu \lambda
^\alpha \,,\nonumber\\
\delta \psi _{\mu L}  & = &
\left( \partial _\mu  +\ft14 \omega _\mu {}^{ab}(e)\gamma _{ab}
+\ft 12\rmi A_\mu^B \right)\epsilon_L +\ft12 M_P^{-2}m\gamma _\mu
\epsilon _R\,, \nonumber\\
\delta \chi _i& = & \ft12\not\! \hat{\partial }\phi_i \epsilon _R- \ft12
(g^{-1})_i^j m_j\epsilon _L\,,
 \nonumber\\
\delta \lambda^\alpha  &=&\ft14\gamma ^{\mu \nu } F_{\mu \nu
}^\alpha\epsilon +\ft12\rmi \gamma _5 (\Re f)^{-1\,\alpha \beta}{\cal P}
_\beta \epsilon  \,. \label{gfsusyphenom}
\end{eqnarray}

Here
\begin{equation}
  A_\mu ^B=\frac{1}{2 M_P^2}\rmi\left[ (\partial _i K)\partial _\mu \phi^i-(\partial
  ^i K )\partial _\mu \phi_i\right]
  +\frac{1}{M_P^2}W_\mu ^\alpha {\cal P}_\alpha \,.
  \label{U1connection}
\end{equation}

It is instructive to rewrite this expression for the composite gauge
field  $A_\mu ^B$ in a slightly different form, where we split the ${\cal
P}_\alpha$-term in two parts, as shown in (\ref{Pnew}). The first part
turns the derivatives of the fields $\phi^i$ and $\phi_i$ into covariant
derivatives, the second part contains the FI terms   so that in the
simple case that $\tilde r_\alpha (\phi) = \rmi g\xi_\alpha / 3$ we find
\begin{equation}
  A_\mu ^B=\frac{1}{2 M_P^2}\rmi\left[ (\partial _i K)\hat{\partial }_\mu
\phi ^i-(\partial
  ^i K )\hat{\partial} _\mu  \phi_i\right]
  +\frac{g}{M_P^2}W_\mu ^\alpha {\xi}_\alpha \, ,
  \label{U1connectionNew}
\end{equation}
where $\hat{\partial }_\mu \phi ^i = \partial_\mu \phi ^i -M_P \eta
_\alpha {}^i\, W_\mu^\alpha $.

\subsection{Summary on $R$-symmetry, FI term and superpotential}
\label{ss:summRFIW}

In summary, there are 4 types of contributions to the local $R$-symmetry
connection
\begin{enumerate}
  \item \label{scterms} In the superconformal algebra there is a $U(1)$ that acts as $R$-symmetry.
  Its gauge vector is an auxiliary field, i.e.\ it becomes a composite once the field
  equations are used. Even when there are no other gauge vectors present,
  this $U(1)$ couples to the gravitino.
  This gives rise to the first terms in~(\ref{AmuBinz}) or~(\ref{U1connection}).
  \item \label{covariantizationterms} When we gauge further symmetries by vector multiplets, new terms
  contribute to the $R$-symmetry gauge field. These are the
  covariantizations of the terms in~(\ref{scterms}). They contribute to ${\cal
  P}_\alpha $ as the first two terms of the second line
  in~(\ref{Pgeneral}). These will be important for section~\ref{ss:FIfromanomU1}. Their
vacuum expectation value will be different from zero only if a field
'shifts' under the symmetry, i.e.\ the vacuum expectation value (vev) of
the transformation law is non-zero.
  \item \label{Kanoninvterms} If the symmetries gauged by vector multiplets do not leave the K{\"a}hler potential
invariant, but give rise to a K{\"a}hler transformation as
in~(\ref{deltaalphaK}) depending on the real part of a holomorphic
function $r_\alpha (z)$, then the latter contribute to the connection
$A_\mu ^B$ via additions to ${\cal P}_\alpha $. This is the part that
does \emph{not} appear in the `simplified' case in
section~\ref{ss:simplAction}.
  \item \label{xiterms} Finally, as the previous part defines $r_\alpha (z)$ only up to
  an imaginary constant, one may add a constant, that has been called $\xi_\alpha$,
  again contributing to ${\cal P}_\alpha $ and hence to $A_\mu ^B$. This term is shown
  explicitly\footnote{There is a typo  in an analogous equation in \cite{Wess:1992cp} where the constant FI term is omitted in the gravitino supersymmetry transformations.} in (\ref{U1connectionNew}). It plays a crucial role in establishing an unbroken supersymmetry of $D$-term strings, as shown
  in~\cite{Dvali:2003zh}.
\end{enumerate}
These four types of terms contribute to the $R$-symmetry $U(1)$
connection, but as the first type of terms do not contribute to ${\cal
P}_\alpha $, only the terms of the
types~\ref{covariantizationterms}--\ref{xiterms} contribute to the FI
term, which is the vev of ${\cal P}_\alpha^2$.

Furthermore, the terms of type~\ref{covariantizationterms} do not
contribute to $r_\alpha (z)$, while the requirement on the
superpotential, (\ref{gtransfW}), only depends on $r_\alpha (z)$. Thus,
in case that we have only terms of type~\ref{covariantizationterms}, the
superpotential has to be invariant. A generalization is that both the
K{\"a}hler potential and superpotential are not invariant, but transform
under a simultaneous K{\"a}hler transformation. That is
case~\ref{Kanoninvterms}.

If the K{\"a}hler potential is invariant (`simplified case'), the requirement
on the superpotential only depends on $\xi _\alpha $ (terms of
type~\ref{xiterms}) as shown in~(\ref{RinvariantW}).

The final gauged $U(1)$ symmetries are the linear combinations of the
$U(1)$ of the superconformal group and the $U(1)$ of vector multiplets
that preserve the gauge choice~(\ref{U1gauge2}). Thus, if $r_\alpha(z)=0$
the symmetry is preserved without corrections of the superconformal
$U(1)$. If, on the other hand, $Y$ transforms under a $U(1)$, i.e.\
$r_\alpha (z)\neq 0$, then the gauge symmetry in the Poincar{\'e} theory is a
mixing of the gauge symmetry with the superconformal $U(1)$ and hence the
gravitino transforms under this symmetry. Thus, this mixing appears
through terms of type~\ref{Kanoninvterms} and~\ref{xiterms}.

Note also that the distinction between the different terms is dependent
on the K{\"a}hler gauge choice. This is easily understood in the conformal
framework. Remember that the significance of the value of $r_\alpha $ is
the transformation of the conformon field $Y$, see~(\ref{delarhoz}). On
the other hand, K{\"a}hler transformations are redefinitions of the conformon
field by holomorphic functions of the scalars as shown
in~(\ref{firstKatr}). Consider as an example a field $\Phi $ with $\delta
_\alpha \Phi=\rmi $. We can then consider a K{\"a}hler transformation
$Y'=Y\exp (c\Phi)$  for a constant $c$, which changes the K{\"a}hler
potential to ${\cal K}'={\cal K}+3 c(\Phi +\Phi ^*)$. Then the new
conformon transforms under the $U(1)$ as $\delta _\alpha Y'=\rmi c Y'$,
or $r_\alpha =\rmi c$. The change in K{\"a}hler potential changes the
contributions to ${\cal P}_\alpha $ of type~\ref{covariantizationterms},
but this is exactly compensated by the fact that the new $r_\alpha $
means an explicit FI term (type~\ref{xiterms}) with $g\xi M_P^{-2}= 3c$.
We thus see that the distinction between the terms depends on the K{\"a}hler
gauge, but this gauge choice then has repercussions on the
superpotential.

Briefly summarizing, there are two cases to be distinguished for the
superpotential. First is the case with constant FI term, where the
superpotential transforms as (\ref{gtransfW}). The second case is when an
effective FI term is produced as the vev of a field that shifts under
$U(1)$. In this case, the superpotential is invariant
(case~\ref{covariantizationterms} above), unless the symmetry only
preserves the K{\"a}hler potential up to a K{\"a}hler transformation, in which
case the superpotential should also have a homogeneous K{\"a}hler
transformation (case~\ref{Kanoninvterms}). The fact that the shift of a
field is promoted into an invariance group of the superpotential, implies
important constraints on its form, as discussed in different parts of our
paper. The question we wish to briefly address now is the interpolation
between the two regimes. In a theory with several chiral fields, one
would expect that if we ``freeze'' the vev of one of the shifting fields,
the resulting structure of the remaining theory must be such as in the
situation with a constant FI term. For instance, we should be able to
define an effective superpotential which starts transforming
non-trivially. In the following section, we shall discuss this connection
in an explicit example.

\subsection{Supersymmetrically removing chiral multiplets}
\label{ss:susyRemoveCh}

Some chiral multiplets may be massive, and it can be interesting to
consider the theory for fixed values of their complex scalars (and zero
fermions).\footnote{Freezing the vev's of the scalars that shift under
the $U(1)$ symmetry, and integrating them out is a rather delicate thing.
We are interested in the situation when the scalar vev that sets the FI
term is stabilized in such a way that the effect of the FI term on the
remaining fields is non-vanishing. Such a situation can be achieved only
for certain ranges of parameters as we discuss in
section~\ref{ss:FIfromanomU1}.} 

We consider here the case that the K{\"a}hler potential splits between the
heavy fields (denoted by $\rho $) and the light fields (denoted by $S$)
\begin{equation}
  K (\rho ,S,\rho ^*,S^*)= K_{(\rho )}(\rho ,\rho ^*) + K_{(S)} (S,S^*)\,.
 \label{Ksplit}
\end{equation}
We may have several fields of each kind. We continue here with one light
field $S$ and one heavy $\rho $, but the generalization is
straightforward. The superpotential should satisfy~(\ref{gtransfW}), or
explicitly, after the rescaling~(\ref{replacements}),
\begin{equation}
\delta _\alpha W=  M_P\eta _{\alpha \rho }\partial ^\rho W + M_P\eta
_{\alpha S}\partial ^S W=
  -3 r_\alpha W\,.
 \label{Winvsplit}
\end{equation}
The Killing potentials are
\begin{equation}
  {\cal P}_\alpha =\rmi M_P\eta _{\alpha \rho  }\partial ^\rho K+
  \rmi M_P\eta _{\alpha S}\partial ^SK -3\rmi M_P^2r_\alpha\,,
 \label{KillPotSplit}
\end{equation}

Now we want to remove $S$ in a supersymmetric way.
Considering~(\ref{gfsusyphenom}), this implies that the stabilized value
of $\rho $, which we indicate as $\rho _0$  should be such
that\footnote{It is here that we use that the metric is diagonal between
the left and right fields.}
\begin{equation}
\left.m^\rho\right|_{\rho =\rho _0} =\left.{\cal D}^\rho  W\right|_{\rho
=\rho _0}=0\,,\qquad \rightarrow\qquad \left.\frac{\partial ^\rho
W}{W}\right|_{\rho =\rho _0} =-\left.M_P^{-2}\partial ^\rho K_{(\rho
)}\right|_{\rho =\rho _0}\,.
 \label{susyintegration}
\end{equation}
Note that the right-hand side does not depend on $S$. This is the case
for superpotentials of the form
\begin{equation}
  W(\rho ,S)= w_{(\rho )}(\rho )w_{(S)}(S)\,.
 \label{Wsplit}
\end{equation}
(if the condition~\ref{susyintegration} would be valid for all $\rho $,
then the potential should be of this form).  For these superpotentials,
the invariance condition~(\ref{Winvsplit}) is
\begin{equation}
  \frac{\delta _\alpha w_{(\rho )}}{w_{(\rho )} } (\rho )
   +\frac{\delta _\alpha w_{(S)}}{w_{(S)} } (S
  )= -3r_\alpha (\rho ,S)\,.
 \label{WinvrhoS}
\end{equation}

The condition~(\ref{susyintegration}) implies that the terms with $\rho
$-derivatives in both~(\ref{Winvsplit}) and~(\ref{KillPotSplit}) can be
omitted if we use instead of $r_\alpha $
\begin{equation}
  r'_\alpha(S) = r_\alpha +\left. M_P\eta _{\alpha \rho }\frac{\partial ^\rho
  W}{3W}\right|_{\rho =\rho _0}=\ r_\alpha -\left. M_P^{-1}\eta _{\alpha \rho }\partial ^\rho
 K_{(\rho)}\right|_{\rho =\rho _0}\,.
 \label{rprime}
\end{equation}
In the case of~(\ref{Wsplit}) we thus obtain
\begin{equation}
  r'_\alpha (S)= r_\alpha (\rho _0,S) + \frac{\delta _\alpha w_{(\rho )}}{3w_{(\rho
  )}}(\rho _0)=-\frac{\delta _\alpha w_{(S)}}{3w_{(S) }} (S )\,.
 \label{rprimesplit}
\end{equation}
E.g. if the original K{\"a}hler potential and superpotential were invariant,
$r_\alpha =0$ in~(\ref{WinvrhoS}), but the two terms of that equation are
opposite constants, then the final $r'_\alpha $ is a constant. In the
case that the separate K{\"a}hler potentials are invariant, this is an
imaginary constant, the FI-term. Thus here we see how the field-dependent
first term of~(\ref{KillPotSplit}) gives rise to a constant FI term,
\begin{equation}
  g\xi =-\rmi M_P^2\frac{\delta _\alpha w_{(\rho )}}{3w_{(\rho)}}(\rho
  _0)\,,
 \label{effectivexi}
\end{equation}
in the effective model at constant value of $\rho $.

We obtain the same potential if we use the new superpotential
\begin{equation}
  W'(S)= \left.W \rme ^{K_{(\rho )}/(2M_P^2)}\right|_{\rho =\rho _0}\,,
 \label{Wprime}
\end{equation}
which for~(\ref{Wsplit}) is proportional to $w_{(S)}(S)$.


We may also compare with the situation in $N=2$ supergravity. Then
arbitrary FI terms are constant values of the triholomorphic moment maps.
Such constant terms are only allowed when there are no physical
hypermultiplets. In presence of hypermultiplets, the triholomorphic
moment maps are functions of the hyperscalars. Fixing the hypermultiplets
to a constant value is then consistent with a constant FI term for the
effective special K{\"a}hler model of only vector multiplets.

\section{$D$-term inflation revisited } \label{s:DtermcstFI}

We first discuss the standard $D$-term inflation model with constant FI
terms taking into account the fact that the superpotential transforms as
in (\ref{transfW}). We thus consider the simplest model :
\begin{eqnarray}
K &=& \left| \phi_0 \right|^2 + \left| \phi_+ \right|^2 + \left| \phi_-
\right|^2 \ , \nonumber \\
W &=& \lambda \ \phi_0 \phi_{+} \phi_{-} \ , \label{Dinflpot} \\
f_{\alpha \beta} &=& \delta_{\alpha \beta} \ , \nonumber
\end{eqnarray}
where the fields $\phi_0$,  $\phi_+$, $\phi_-$ in the globally
supersymmetric theory have charges ($Q_\pm = \pm 1$, $Q_0 =0$). To
promote this model to local supersymmetry, i. e. to supergravity with
constant FI term, we have to change the charge assignments for the chiral
superfields, so that the superpotential transforms under local
$R$-symmetry. We choose
\begin{equation}
\label{charge} q_i = Q_i - \rho_i {\xi \over M_P^2} \ , \quad
\sum_{i=\pm,0} \rho_i = 1 \,.
\end{equation}
The last equation follows from  (\ref{sumrule}).

The complete scalar potential reads, following (\ref{Vtotal}),
\begin{eqnarray}
V &=& V_F + V_D \ , \nonumber \\
V_F &=& e^{(|\phi_0|^2 + |\phi_+|^2 + |\phi_-|^2)/M_P^2} \left\{ \lambda^2
|\phi_+|^2|\phi_-|^2 \left( 1 + {|\phi_0|^4 \over M_P^4}\right) +  \lambda^2
|\phi_0|^2|\phi_-|^2 \left( 1 + {|\phi_+|^4 \over M_P^4}\right) \right.
\nonumber \\
&& \qquad \qquad \qquad \qquad \left. + \lambda^2 |\phi_0|^2|\phi_+|^2
\left( 1 + {|\phi_-|^4 \over M_P^4}\right) + 6 \lambda^2
{|\phi_0\phi_+\phi_-|^2 \over M_P^2} \right\}, \label{DinflVF} \\
V_D &=& {g^2 \over 2} \left[ q_0 |\phi_0|^2 + q_+ |\phi_+|^2 + q_-
|\phi_-|^2 - \xi \right]^2 \ . \label{DinflVF2}
\end{eqnarray}
For fixed $\phi_0$, we have
\begin{equation}
V = {g^2 \over 2} \left[q_0 |\phi_0|^2 - \xi \right]^2 + \sum_{i=\pm}
|\phi_i|^2 \left[ (\lambda^2 \rme^{|\phi_0|^2 /M_P^2} + g^2
q_iq_0)|\phi_0|^2 -
 g^2\xi q_i \right] + {\cal O}(\phi_\pm^4)\,.
\end{equation}
Hence for $|\phi_0|^2 >  g^2\xi q_\pm/(\lambda^2 e^{|\phi_0|^2 /M_P^2}+
g^2 q_0q_\pm)$, the minimum is found for $\phi_\pm = 0$, direction along
which
\begin{equation}
V(\phi_0, \phi_\pm=0) = {g^2 \over 2} \left[q_0 |\phi_0|^2 - \xi
\right]^2 \ .
\end{equation}
Since $q_0 = -\rho_0 \xi/M_P^2$, we conclude that for $\rho_0 \not = 0$,
the mass of the neutral scalar is given by $m_{\phi_0}^2 \sim
g^2\xi^2/M_P^2 \sim H^2$. To recover the properties that motivated
$D$-term inflation we must require $\rho_0 = 0$, i.e.
\begin{equation}
q_0 = 0 \ , \label{q0}
\end{equation}
in which case the potential has a plateau at $V_0 =  g^2\xi^2/2$.

We now turn to the evaluation of the one-loop potential along this flat
direction. The corresponding scalar and fermion masses are
\begin{eqnarray}
m^2_{\phi_\pm} &=& \lambda^2e^{|\phi_0|^2 /M_P^2} |\phi_0|^2 - g^2 \xi q_\pm \ , \nonumber \\
 m^2_{\chi_\pm} &=& \lambda^2 |\phi_0|^2e^{|\phi_0|^2 /M_P^2}   \ .
\end{eqnarray}
In fact, the quantities in~(\ref{fermionmasses}) are all zero at
$\phi_\pm=q_0=0$ except for
\begin{equation}
  m^{+-}=m^{-+}=\lambda \phi _0e^{|\phi_0|^2 /(2M_P^2)}\,.
 \label{m+-}
\end{equation}
The contribution to the effective potential is proportional to $\STr  M^4
\ln (M^2 / \Lambda^2)$. We neglect the $ \xi q_\pm $ term in the $\ln$
factor, and the $\rme^{|\phi_0|^2 /M_P^2} $ factor that went with
$\lambda ^2$. Hence
\begin{equation}
\STr  M^4 \ln {M^2 \over \Lambda^2} = \left[ -2 g^2\xi \lambda^2
|\phi_0|^2 (q_+ + q_-) +  g^4\xi^2 (q_+^2 + q_-^2) \right] \left( \ln
{\lambda^2 |\phi_0|^2 \over \Lambda^2}+\frac{|\phi_0|^2}{M_P^2}\right),
\end{equation}
where, following (\ref{charge}) and (\ref{q0}), $q_+ + q_- = -\xi/M_P^2$
and $q_+^2 + q_-^2 \sim 2$. Thus
\begin{eqnarray}
V(\phi_0, \phi_\pm=0) &=& V_0 + {1 \over 64 \pi^2} \ \STr M^4 \ln {M^2
\over \Lambda^2} \nonumber \\
&=& {g^2 \over 2}  \xi^2 \left[ 1 + {1 \over 16 \pi^2} \left(1 +
\lambda^2 {|\phi_0|^2 \over M_P^2}\right) \ln {\lambda^2 |\phi_0|^2 \over
\Lambda^2} \right] \ . \label{V1loop}
\end{eqnarray}

Next let us discuss the problem of anomalies. The charge of the fermion
$\chi_i$ is the strength of the coupling to $W_\mu$ in the covariant
derivative~(\ref{Rchiralf}). This can be written as
\begin{equation}
  {\cal D}_\mu \chi _i=\left( \partial _\mu  +\ft14\omega _\mu
  {}^{ab}(e)\gamma _{ab}+\ft 14M_P^{-2}
  (\phi _j\partial _\mu \phi ^j -\phi ^j\partial _\mu \phi_j)
   -\rmi g\tilde q_i W _\mu \gamma _5
 \right) \chi _i \,.
 \label{introtildeq}
\end{equation}
Hence the charge is
\begin{equation}
\tilde q_i = q_i + G = q_i + {1 \over 2} {\xi \over M_P^2} \,.
\label{tildeq}
\end{equation}
Similarly, (\ref{Rcharges}) yields $\tilde q_{\lambda} =-G$ for the
gaugino and $\tilde q_{\psi} = -G$ for the gravitino.

Then the $U(1)^3$ anomaly coefficient $C$ reads
\begin{equation}
C = \sum_{i=0,\pm} \left( q_i + G\right)^3 + (-G)^3 + 3 (-G)^3 \,,
\label{C}
\end{equation}
where the last two terms are respectively the gaugino and gravitino
contributions. Using $\rho_+ + \rho_-=1$, we have $q_+ + G = 1 + (1-2\rho_+)G
= 1 - (1-2\rho_-)G = -(q_- +G)$ and the anomaly coefficient reduces to the
gravitino contribution: $C=-3G^3$.

Adding 3 neutral ($q=0$, hence $\tilde q=G$) chiral multiplets can cancel
these anomalies. These extra fields should not occur in the
superpotential $W$. This gives a generalization of the $D$-term inflation
model taking into account the terms in supergravity that provide an exact
local supersymmetry of the classical action.

In the theories where the gauge anomaly is not cancelled (anomalous
$U(1)$) it is possible to introduce a coupling to the axion $a$ of the
type $aFF^*$. The shift of the axion field under $U(1)$ may remove  the
anomaly. However, this requires to introduce a coupling of the form $(\Re
f(z)) F^2+ (\Im f(z)) FF^*$ and stabilize the additional field $z$. We
will discuss this case in section~\ref{ss:FIfromanomU1}. But  here, since
we have a cancellation of all $FF^*$, the $D$-term inflation model with
constant FI term, supplemented by 3 neutral chiral multiplets, is valid
in the original version with constant kinetic function for the vector
multiplet.

For the  gravitational anomaly we find:
\begin{equation}
C_g = \sum_{i=0,\pm} \left( q_i + G\right) + (-G) - 21 (-G) +3G = 24 G
\,. \label{Cg}
\end{equation}
The 3 extra chiral multiplets introduced to cancel the gauge anomalies
cannot cancel the gravitational ones. The relevant terms of the form
$RR^*$ are higher-derivative terms which are not present in the classical
supergravity action. One can think that the anomaly terms of this kind
should be taken care in the context of other higher-derivative terms in
the action and most likely in the context of the full M/string theory.

Our observation about  neutral chiral multiplets cancelling the gauge
anomaly of gaugino and gravitino can be applied to the Freedman
model~\cite{Freedman:1976uk}. When it is supplemented by 4 neutral chiral
multiplets, they cancel the $FF^*$ anomaly.

\section{$D$-term strings}
\label{s:DtermStrings}

In theories with FI $D$-term inflation, the $U(1)$ symmetry gets
spontaneously broken. This breaking in general can result into the
formation of the cosmic strings, as explained in~\cite{Dvali:2003zh},
where these were referred to as $D$-term strings. It was shown that the
$D$-term strings are (the only) BPS-saturated strings in $N=1$, $4d$
supergravity, and therefore are the natural candidates for the low energy
description of $D$-brane strings.  We shall postpone exploration of this
connection to sections~\ref{s:cosmoApplicD}-\ref{ss:FIDfromD}, and here
we will study some unusual properties of these objects within $4d$
supergravity.

\subsection{The string configuration}
\label{ss:Dstringconfig}

The string configuration can be obtained from one vector multiplet with
minimal kinetic term, and one chiral multiplet, charged under the $U(1)$
of the former with charge $q=1$ and with minimal K{\"a}hler potential $K=\phi
\phi ^*$. The solution that we consider is purely bosonic. The scalar of
the chiral multiplet depends only on two coordinates of the $3+1$
dimensional space, which are parametrized by a distance $r$ from the
string, and  an azimuthal angle $\theta$, and has the form
\begin{equation}
\phi (r,\theta )\, = \, f(r)\,{\rm e}^{\rmi n \theta}\,.
 \label{stringhiggs}
\end{equation}
$f(r)$ is a real function that outside the string core approaches the
vacuum value $f^2=\xi$.  The gauge potential takes the form
\begin{eqnarray}
 g W_\mu\, \rmd x^\mu = n\alpha (r) \,\rmd\theta \,
  \qquad \rightarrow \qquad
  F=\ft12F_{\mu \nu }\,\rmd x^\mu \,\rmd x^\nu ={n \alpha '(r)\over g} \rmd
  r\,\rmd\theta = {n \alpha '(r)\over  g C(r)} e^1 e^2 \,,
 \label{explicitW}
\end{eqnarray}
where we already used vierbein forms
\begin{equation}
  e^1=\rmd r\,, \qquad e^2=C( r)\rmd\theta \,.
 \label{vierbeine1e2}
\end{equation}
They live in a space which can be described by a metric
\begin{equation}
  \rmd s^2= -\rmd t^2 +\rmd z^2+\rmd r^2 + C^2(r) \rmd \theta ^2\,,
 \label{tentativemetric}
\end{equation}
 which leads to the spin connections
\begin{equation}
  \omega_r{}^{12}=0\,,\qquad
  \omega_\theta {}^{12}= -C'(r)\,.
 \label{12components}
\end{equation}
This defines a BPS configuration if the following differential equations
are satisfied:
\begin{eqnarray}
    C (r)f'(r)&=& |n| f(r)\left[ 1-\alpha (r)\right], \nonumber\\
    \frac{\alpha '(r)}{g C (r)}&=& \frac{g}{|n|} \left[ \xi -
    f^2(r)\right], \nonumber\\
  1-C'(r) &=& \pm  A_\theta ^B=\frac{|n|}{M_P^2}\left[ \xi \alpha (r)-\frac{C (r)}{2 g^2}
  \left( \frac{\alpha '}{C (r)}\right)'\right]\,,
 \label{explicitBPSrho}
\end{eqnarray}
where $\pm =n/|n|$. Then there is a residual supersymmetry
\begin{equation}
  \epsilon =\rme^{\mp \rmi\theta \gamma _5 /2} \ft12\left( \unity \pm \rmi \gamma _5 \gamma ^{12}\right)
\epsilon _{0} \label{preservedSusy}
\end{equation}
where $\epsilon _0$ is a constant spinor, of which the previous factor,
$\Pi _\pm$ in the terminology of~(\ref{Pipm}), selects 2 independent real
components. In rigid supersymmetry this preservation of 1/2 of
supersymmetry was found in~\cite{Dvali:1997bg,Davis:1997bs}. In
supergravity the BPS condition on the gravitino is also satisfied due to
a conspiracy of the spin connection~(\ref{12components}) and the
$R$-symmetry connection~(\ref{U1connectionNew}) similar to the mechanisms
in $2+1$ dimensions~\cite{Becker:1995sp,Edelstein:1996ba}.

We thus find that the $D$-term strings with elementary flux are
BPS-saturated states, and preserve half of the supersymmetry.

\subsection{Zero modes on D-strings} \label{s:zeromodesD}

We have shown that D-strings are BPS saturated objects and preserve one
half of the original supersymmetry. In view of this fact, the zero modes
on the string exhibit a somewhat puzzling behaviour.
 To see this, let us complicate the model a bit in the following way.
We shall couple the Higgs field $\phi$, that forms a D-string, to some
number of chiral superfields $\Phi_{i}$ in the superpotential
\begin{equation}
W \, = \, {a_{i} \over 2} \phi\, \Phi_{i}^2\,, \label{species}
\end{equation}
where $a_{i}$ are coupling constants. We shall denote the fermionic
components of these superfields by $\chi_{i}$, and the scalar component
by $\Phi_{i}$. For simplicity, let us consider a single species of such
fermions, the generalization to an arbitrary number of species being
trivial. Also, the puzzle that we wish to discuss already appears in the
limit of rigid supersymmetry, and switching on supergravity does not
change it much. So for the beginning let us discuss the issue in this
limit. In the case of rigid supersymmetry, the $U(1)$ is not an
$R$-symmetry and the charges of the scalar $\Phi$ and fermion $\chi$ are
equal. This charge (call it $q_{\chi}$) must be exactly $-{1\over 2}$ of
the $\phi$ charge (which we normalize to one), and thus contributes to
${\cal P}$ with the same sign as the FI term (taking $\xi >0$). This
implies that the expectation value of the scalar $\Phi$ is identically
zero everywhere in the string background, and our $D$-string solution is
unaffected. Indeed, the mass-square of the $\Phi$-scalar is
\begin{equation}
m^2_{\Phi} \, = \, |a \phi|^2 \, + \, \ft12g^2 \left[ - |\phi|^2 + \xi
\right]. \label{massofphi}
\end{equation}
This quantity is positive definite everywhere (for $|a|^2\geq\ft12g^2$),
including the string core, and hence the lowest energy configuration
implies $\Phi \, = \, 0$. Hence, even in the presence of the
$\Phi_{i}$-fields, the string background preserves half of the
supersymmetry.

We shall now discuss the zero modes in the string background. Notice that
the phase of the mass of the fermion $\chi$ changes by $2\pi$ around the
strings and therefore according to standard index
theorems~\cite{Weinberg:1981eu,Jackiw:1981ee} there must be a
normalizable fermionic zero mode trapped in the core of the string. Let
us find this mode explicitly. We put $\Phi =0$ and take the
configuration~(\ref{stringhiggs}) with $n=1$.

The Dirac equation for $\chi$ in the string background is
\begin{equation}
\not\!\!
{\cal D} \chi_L \, = -\, a f(r) {\rm e}^{-\rmi \theta} \,
\chi_R\,. \label{dirac}
\end{equation}
After standard separation of variables by transverse and longitudinal
functions,
\begin{equation}
  \chi \, = \, \alpha(t,z)\, \tilde{\chi}(r, \theta)\,,
 \label{sepvariables}
\end{equation}
we arrive to the following Dirac equation for the spinor $\tilde{\chi}$
\begin{equation}
\left [ \gamma^1\,  \partial_r \, + {\gamma^2 \over C(r)} \,
\left(\partial_{\theta}\,-\ft12\gamma _{12}C'(r)  - \, \rmi \left(
gq_{\chi}\, W_{\theta}+\ft12A_\theta ^B\right) \right) \, \right ] \,
\tilde{\chi}_L \, = \,- a f(r) {\rm e}^{-\rmi\theta} \, \tilde{\chi}_R\,,
\label{Cdiractr}
\end{equation}
where the term $-\ft12\gamma _{12}C'$ had to be added to the $\theta
$-derivative as spin-connection term in the curved basis $(r,\theta )$.
We then look for a solution\footnote{The $\chi _0$ spinor is in fact the
spinor in a flat basis.}
\begin{equation}
  \tilde \chi (r,\theta )= \rme^{\gamma _{12}\theta /2} \chi _0(r)\,,
 \label{tildechichi0}
\end{equation}
such that~(\ref{Cdiractr}) reduces to
\begin{eqnarray}
&&\gamma^1\left [ \,  \partial_r \, + { 1\over C(r)} \,\left(
-\ft12+\ft12C'-\rmi\gamma^{12}\left(g q_{\chi}\, W_{\theta}+\ft12A_\theta
^B\right)\right) \right ] \,\rme^{\gamma _{12}\theta /2} \chi _{0L}(r)
\nonumber\\
&& = \,- a f(r) {\rm e}^{-\rmi\theta} \, \rme^{\gamma _{12}\theta/2 }
\chi _{0R}(r)\,. \label{Cdiracr}
\end{eqnarray}
Using projected spinors (see appendix~\ref{app:decompsusy})
\begin{equation}
  \gamma _{12}\chi _L^\pm = \mp \rmi \chi_L^\pm\,,\qquad
\gamma _{12}\chi _R^\pm = \pm \rmi \chi_R^\pm\,,
 \label{chi+-}
\end{equation}
and the third of~(\ref{explicitBPSrho}) we find
\begin{eqnarray}
&&\gamma^1\left [ \,  \partial_r \,+ { 1\over C(r)} \,\left(
-\ft12A_\theta ^B-\rmi\gamma^{12} \left( g q_{\chi}\,
W_{\theta}+\ft12A_\theta ^B\right)\right) \right ]
\left(\rme^{-\rmi\theta /2} \chi^+ _{0L}(r) +\rme^{\rmi\theta /2}\chi^-
_{0L}(r)\right)
\nonumber\\
& &= \,- a f(r) \left(\rme^{-\rmi\theta/2 }  \chi^+ _{0R}(r)
+\rme^{-3\rmi\theta/2 } \chi^- _{0R}(r)\right)\,. \label{Cdiracrproj}
\end{eqnarray}
Therefore only the $+$ modes exist, for which we get the relation
\begin{equation}
\gamma^1\left [ \,  \partial_r \, - {1 \over C(r)} \left( gq_{\chi}\,
W_{\theta}+A_\theta ^B\right)\right ]  \chi^+ _{0}(r)
 = \,- a f(r)  \chi^+ _{0}(r) \,. \label{Cdiracr+}
\end{equation}

The solution is thus given by two modes, decomposing the $+$ mode using
further projections $(\unity \pm \gamma _1)$
\begin{eqnarray}
  \chi^+ _{0}(r)&=& \rme^{\int_0^r
  \left[ \left(
gq_{\chi}\, W_{\theta}+A_\theta ^B\right)(r') /C(r') - a f(r')\right]\rmd
r' }(\unity +\gamma _1)
  \chi_0^+ +\nonumber\\
  &+& \rme^{\int_0^r
  \left[ \left(
gq_{\chi}\, W_{\theta}+A_\theta ^B\right)(r') / C(r') + a
f(r')\right]\rmd r' }(\unity -\gamma _1) \chi
  _0^+\,.
 \label{Csolnchi}
\end{eqnarray}
The constants $(\unity \pm \gamma _1) \chi_0^+$ are two real modes of
fermions.

Since $W_\theta (r)$ goes to a constant, $C(r)$ is linear in $r$ in the
rigid limit and $f(r)$ approaches the constant $\sqrt{\xi}$ at infinity,
one of the zero modes is normalizable. Hence, for the string with unit
winding number, there is one normalizable fermionic zero mode in the
spectrum of string excitations. For $n$ windings the number of zero mode
solutions is $n$ according to the index theorem. This result can be
trivially generalized to an arbitrary number of $\Phi_i$ superfields. In
the background of the string with winding number $n$, each $\chi_{i}$
fermion deposits $n$ normalizable zero modes. Hence, we can arbitrarily
increase the number of the fermionic zero modes, either by increasing the
winding number or the number of chiral fermions species coupled to the
Higgs field.

 Taking into the account the fact that the D-string leaves half of the
supersymmetry unbroken, we might have expected to find an equal number of
bosonic zero modes, coming from the $\Phi_{i}$ fields. However, we find
none! In the $\Phi_{i}$ spectrum, the localized scalar excitations only
exist if $ a \sim g$ or larger (whereas localized fermionic zero modes
exist for any $a\neq 0$), and even in this case the masses of the lowest
scalar excitations are $\sim \xi$. Thus, the number of bosonic and
fermionic zero modes is clearly unbalanced.


The resolution of the puzzle is that the unbroken supersymmetry acts on
the fermionic zero modes trivially, and therefore there are no bosonic
partners. In other words, whereas in the usual case we would obtain the
bosonic partner by super-shifting the fermion, here the shift of the zero
mode fermion vanishes
\begin{equation}
 \Phi_i(\epsilon ) = \bar \epsilon_L\Delta \chi _i\, = 0\,, \label{zeroshift}
\end{equation}
(referring to the notations in appendix~\ref{app:BosFerModes}). Indeed,
the transformation parameters of unbroken supersymmetry and the zero
modes are both $+$ spinors under the projections~(\ref{Pipm}), and the
property~(\ref{barpm}) then kills the bosonic mode~(\ref{zeroshift}).

This can be understood from the algebra of the remaining supersymmetry.
In fact, using the same type of projections, the non-zero part of the
algebra is only non-zero between two supersymmetries of opposite type,
see~(\ref{Q+Q-}). Therefore the preserved supersymmetry is nilpotent. An
equal number of bosonic and fermionic modes is only expected when the
supersymmetry squares to an invertible operator (e.g. the translations).

We would like to note that in \cite{Becker:1995sp} (realizing an
observation by Witten~\cite{Witten:1995cg}) a $2+1$-dimensional BPS
vortex solution without Fermi-Bose degeneracy was found. Our effect,
however, is of the different origin, since, unlike
in\cite{Becker:1995sp}, in our case the Fermi-Bose non-degeneracy
persists even in the globally-supersymmetric limit.


%
%

\section{Discussion of FI term from pseudo-anomalous $U(1)$.}
\label{ss:FIfromanomU1}

We now turn to the case of field-dependent FI terms. As is well-known,
such field-dependent FI-terms emerge whenever there is a chiral
superfield, which we denote $\Phi$, shifting under $U(1)$. In particular,
this is mandatory whenever the $U(1)$ symmetry exhibits a chiral anomaly
that is cancelled by the Green-Schwarz mechanism \cite{Green:1984sg}. In
such a case, the imaginary part of the $\Phi$-scalar plays the role of an
axion, and cancels the chiral anomaly by shifting under the $U(1)$
symmetry. Such a $U(1)$ is sometimes referred  to as  ``anomalous'' or
more appropriately
``pseudo-anomalous'' since the total anomaly is of course zero.
The FI term in such a case depends on the real part of $\Phi$.

The supersymmetric set up for anomalous FI terms was developed in 1987 in
\cite{Dine:1987xk} in the context of the heterotic string theory. It was
used since then in the cosmology literature under the assumption that the
dilaton field $\Phi$ is somehow stabilized and therefore that the
dilaton-dependent $D$-term produces a  constant FI term. Now that we are
trying to understand the $D$-term potentials in supergravity and string
theory at the fundamental level, we have to revisit this approach.

For the purpose of illustration, we consider a ``dilaton field'' $\Phi $,
adding to the K{\"a}hler potential a part $K_{(\Phi )}(\Phi + \bar \Phi )$.
This field transforms under the $U(1)$ as
\begin{equation}
  \delta \Phi =\rmi \alpha \,,\qquad \mbox{or}\qquad \eta _\Phi =\rmi M_P^{-1}\,.
 \label{U1Phi}
\end{equation}
The K{\"a}hler potential is invariant, implying that the real part of $r$ is
zero, and we do not add an imaginary constant to $r$, which would imply
the presence of the constant FI term. However,~(\ref{U1Phi}) gives an
extra contribution to ${\cal P}$:
\begin{equation}
  {\cal P}_{(\Phi )}= -K_{(\Phi )}'(\Phi + \bar \Phi )\,.
 \label{calPPhi}
\end{equation}
If this has a non-zero constant, it acts as a FI term.

Actually in \cite{Dine:1987xk} a ``stringy'' \Ka\ potential is
considered, with appropriate vector kinetic term
\begin{equation}
  K= -\ln(\Phi +\bar \Phi)\,, \qquad  f= \Phi \,, \qquad \Phi \equiv \phi ^{-2}
  + \rmi  b\,.
 \label{stringyK}
\end{equation}
The gauge coupling thus depends on the dilaton, $(\Re f)^{-1}=\phi^2$.
The theory also has an axion coupling proportional to $b\,F\,F^*$. With
the shift transformation of the axion field under $U(1)$ as in
(\ref{U1Phi}), this term serves to remove the anomaly proportional to
$FF^*$.

This gives
\begin{equation}
  {\cal P}_{(\Phi )}= - K_{(\Phi  )}'(\Phi  + \bar \Phi  )= {1\over \Phi  + \bar \Phi }
  =\frac12  \phi^2\,.
 \label{calPPhi2}
\end{equation}
The $D$-term potential is
\begin{equation}
\ft12 (\Re f)^{-1} {\cal P}^2 =\ft18 \phi^6 \,.
\label{heterotic}
\end{equation}
Note that if this would be the only dependence on $\phi^2$, the potential
would tend to make $\phi^2 \rightarrow 0$ and the FI term would
disappear.

The stabilization of the dilaton was assumed in \cite{Dine:1987xk} and
therefore it was considered that  the dilaton-dependent  $D$-term can be
qualified as a constant  FI term. Its value was computed in the context
of the weakly-coupled heterotic string for a constant string coupling
$g_s$ (constant dilaton, $ g_s= \phi^2$) and found to be
\cite{Atick:1987gy}:
\begin{equation}
\xi_{GS}  = {g_s^2 \, \Tr Q\over 192 \pi^2}  M^2_P\, .
\label{anomalousFI}
\end{equation}
The potential  energy is given by $V_D= {1\over 2}g_s^2 \xi_{GS}^2 \sim
g_s^6$. Clearly, without an assumption that $\phi$ is constant, this
potential would behave, as we have just seen, as $ \phi^6$ and tends to
zero at small $\phi$.

Based on this specific analysis, two criticisms were made to the $D$-term
inflation scenario. First (see for example \cite{Dvali:1998mh}), the
inflation scale provided by (\ref{anomalousFI}) is too close to the
Planck scale to be consistent with the COBE normalization. Secondly, the
necessary stabilization of the dilaton field requires some non-vanishing
$F$-terms which, in the context of the simple model described by
(\ref{stringyK}), drown any $D$-term \cite{Arkani-Hamed:2003mz}. We will
discuss this latter question at the end of this section and first address
the former criticism.

Quite generally, the role of the field $\Phi$ may be played by any
modulus, whether it is the dilaton or any volume modulus. The situation
with the stabilization of dilaton and volume in heterotic string theory
is quite complicated. A few recent studies
\cite{Becker:2003sh,Gukov:2003cy,Buchbinder:2003pi} have shown that
stabilization of moduli in heterotic string theory requires significant
deviations from the weakly coupled regime of the heterotic theory, such
as strong coupling, presence of M5-branes etc. Therefore the actual
numbers for FI terms, used in the cosmological context on the basis of
weakly coupled  heterotic string theory, have to be revised. One more
important phenomenological ingredient for the $D$-term inflation model
was the choice of the gauge coupling: it was assumed to be of the same
magnitude as the coupling of the standard model $U(1)$. One may, however,
expect few more $U(1)$ in string theories, some of which may have
different scales of gauge coupling and therefore give different values
for the $D$-term inflation model.

In other string theories, the $D$-term may depend on some other moduli,
like in type IIB theory where it is the volume of the compactified
directions \cite{Burgess:2003ic}. The gauge coupling for the vector field
on the D7 brane has a dependence on the volume of the compactification
modulus: $\Re f\sim (\rho+\bar \rho)$. The \Ka\ potential is  $K=
-3\ln(\rho+\bar \rho)$. As a result, the field-dependent $D$-term related
to the non-self-dual fluxes on D7 brane is given by
\begin{equation}
{1\over 2}(\Re f)^{-1} {\cal P}^2 = {C\over (\rho+\bar \rho)^3}\, .
\label{IIB}
\end{equation}

In both cases the $D$-term potential has a runaway behaviour. Thus, using
these potentials we cannot rely on any assumption of stabilization. To
find models for cosmology from string theory, one has to consider the
total potential where the dilaton and/or the volume have to be
stabilized. Only when this is done, one will be  able  to keep the
version of $D$-term inflation based on anomalous stringy $U(1)$ valid.

In short, the stringy FI term required for GS mechanism of anomaly
cancellation, is a field-dependent $D$-term. Before one stabilizes it, it
cannot be  used  in the cosmological context.  As we already explained in
the Introduction, it is not known how to derive constant FI terms from
string theory, despite the fact that they are allowed under certain
conditions in $N=1$, $d=4$ supergravity.

Let us finally  discuss some generic issues that arise when one tries to
stabilize FI terms in the field-dependent case.

A naive approach would be to assume that the vev of $\Phi$ is stabilized
at some arbitrarily high scale $M_{\rm st}$, by some superpotential, in
such a way that below the scale $M_{\rm st}$ we can integrate out $\Phi$
in order to be left at low energies with a {\it supersymmetric} theory
with a constant FI term. This is, however, not possible as can be seen
from the two following arguments.

 First, if we manage to give a large supersymmetry-preserving
mass to the real part of $\Phi$, then by supersymmetry this must also be
the mass of its imaginary part. However, this imaginary part is the axion
that shifts under $U(1)$, and thus gets a mass through the Higgs effect
by becoming the longitudinal component of the $U(1)$-gauge field. Hence,
if supersymmetry is preserved, $\Phi$ simply becomes a part of a massive
vector supermultiplet with mass $M_{\rm st}$. Hence, $\Phi$ {\it cannot}
be integrated out in a supersymmetric way, unless we integrate out the
whole massive vector supermultiplet. This fact immediately implies that
if we integrate out $\Phi$ below scale $M_{\rm st}$ we cannot be left
with a non-zero FI term. Presence of a non-zero FI term in an effective
supersymmetric theory at any scale requires an existence of a
corresponding vector superfield, unless supersymmetry is broken.

 Hence, $\Phi$ cannot be stabilized at scales bigger than the
effective low energy value of the FI term in a susy-preserving way,
without jeopardizing the very existence of FI in the low-energy theory.
Thus, in general, stabilization of $\Phi$ requires some additional
supersymmetry breaking, e.g. via some non-zero $F$-terms. We will see
below that these $F$-terms can be parametrically (depending on the
parameters in the K{\"a}hler potential) smaller than the FI-terms.

The fact that $\Phi$ cannot be stabilized without the additional
supersymmetry breaking is also clear from the following argument based on
the form of the superpotential. Because $\Phi$ shifts under the gauged
$U(1)$ symmetry, the superpotential must be invariant under this shift
(up to an arbitrary K{\"a}hler transformation, and assuming now that there is
no explicit FI term $\xi $). Hence it can only depend on $\Phi$ through
invariants of the form (or functions of them)
\begin{equation}
  W \, = \, \rme^{\Phi c} w_0
 \label{WeYc}
\end{equation}
 where $c$ is some constant, and $w_0$ is a holomorphic function of
other chiral superfields such that it carries an overall charge $Q = -c$
under the $U(1)$ symmetry. $w_0$ may in particular be a composite
operator, generated by some strong dynamics, such as e.g., gaugino
condensation in a strongly coupled $SU(N)$ group.  In each particular
case $w_0$ will be subject to the $U(1)$-selection rules resulting from
charge assignment dictated by GS anomaly cancellation.

 Thus, at the minimum of the $\Phi$-potential, the $F$-term of $\Phi$ is
generically non-zero. Indeed, it is impossible to satisfy the equations
$D_{\Phi}W =0$, at the true minimum of the potential, because the
$D$-term has no minimum, only a runaway.  So to stabilize $\Phi$, the
$F$-terms should become non-zero. On the other hand, the $F$-terms have
to be smaller than the $D$-terms if we want to have the $D$-term
inflation. This can result from a suitable K{\"a}hler function.\footnote{Its
origin is of course a separate issue, not to be
discussed here.} 
This K{\"a}hler function can only stabilize $\Phi$ if there are some non-zero
$F$-terms. As said above, such terms are expected due to the form of the
stabilizing superpotential, and due to the tendency of the FI term to
push $\Phi$ to the runaway branch.
To have the $D$-term inflation we require that
\begin{equation}
  F \ll D\,,
 \label{FsmallerD}
\end{equation}
 and also we need to satisfy the following inequality:
\begin{equation}
  H^2 \, = \, g^2\xi^2/M_P^2 \,  \ll  \, m^2_{\Phi}  \,  \ll g\xi\,,
 \label{H2orF2}
\end{equation}
where $m_{\Phi}^2$ is the stabilizing mass of $\Phi$. The above implies
that we need the $\Phi$ mass to be bigger than the Hubble parameter
during inflation, which is also of the same order as the contribution
from the $D$-terms into the mass of $\Phi$. In other words, we need $F$
and $D$-terms to split their roles in the following way, that $D$-terms
drive inflation, but $F$-terms stabilize $\Phi$. This puts a requirement
on the K{\"a}hler function.


If a stabilization of the dilaton and volume would be established in some
models, then one would have a mechanism for deriving  $D$-term inflation
from string theory. Recently a proposal for stabilization of the dilaton
and volume modulus was suggested in \cite{Kachru:2003aw} in the context
of the $F$-term potential. Some versions of brane inflation
\cite{Dvali:1998pa}  were studied in \cite{Kachru:2003sx} where
particular difficulties with realization of the brane inflation with
dilaton and volume stabilization were pointed out. More recently,  some
proposals were made \cite{Hsu:2003cy,Firouzjahi:2003zy} towards
improvement of this situation due to a particular shift symmetry of the
potential associated with the BPS states of branes. These proposals  are
supported by the supergravity analysis of type IIB compactifications in
presence of fluxes and branes in \cite{Angelantonj:2003zx,Koyama:2003yc}.
There is a hope that these efforts may lead to a satisfactory derivation
of inflation from string theory.

\section{Cosmological applications: Stability of the $D$-term strings}
\label{s:cosmoApplicD}

 We wish now to briefly discuss some aspects of the cosmological implications of the $D$-term strings.
It is sometimes assumed that  $D$-term strings are necessarily formed at
the end of the $D$-term inflation, and their tension may be in conflict
with the observed spectrum of density perturbations.
 However, this expectation is too naive and in reality the issue is much more subtle.
In brief, we will see that there is no reason to think that the $D$-term
cosmic strings cause any observable problems: in many models D-strings
may not be topologically stable.

More importantly perhaps, it has been conjectured recently
\cite{Dvali:2003zh} that there is a correspondence between the D-strings
of type $II$ string theory, and $D$-term strings. This idea was further
explored in \cite{Halyo:2003uu}. According to this conjecture, BPS
D$_{1+q}$ branes wrapped on a $q$-cycle, are seen from the point of view
of $4d$ supergravity as $D$-term strings. This conjecture has some
immediate implications for the $D$-term and D-brane cosmology. For
instance, since according to current understanding \cite{Sen:1998ii},
D$_{1+q}$-branes can be thought of as the tachyonic vortices formed in
the annihilation of (D$_{3+q} - \bar{\mbox{D}}_{3+q}$)-branes, it then
immediately follows from the conjecture in \cite{Dvali:2003zh}  that the
(D$_{3+q} - \bar{\mbox{D}}_{3+q}$)-system corresponds to a non-zero
$D$-term.\footnote{We learned from J. Maldacena that M. Douglas also
noticed that brane-anti-branes are $D$-terms.} Moreover, if the
compactification volume is somewhat larger than the string scale, the
D-string tension can be easily lowered in order to accommodate the
current observational bounds. Hence, in type $II$ theories the cosmic
D-strings do not cause any cosmological problems and in fact may be
potentially observable in the form of supergravity
$D$-term strings.  

Recently, the suppression of the cosmological production of cosmic
$D$-strings in type $II$ theories was studied in \cite{Dvali:2003zj}.
Some of the interesting potential instabilities of these objects were
pointed out in \cite{Copeland:2003bj}. D-brane--$D$-term-string
correspondence then allows to look for the counterparts of all these
effects in the supergravity $D$-term strings. A number of such
connections were demonstrated in \cite{Dvali:2003zh}. Below we wish to
provide additional links. We shall first discuss instabilities of
$D$-term strings in $4d$ supergravity, and then relate these
instabilities to the ones of $D$-strings in type $II$ theory discussed in
\cite{Copeland:2003bj}.

\subsection{$D$-term string stability in supergravity}
\label{ss:DstabilSG}

 We first  discuss the case of supergravity.
As shown in \cite{Dvali:2003zh}, in the simplest models, in which the
constant FI $D$-term gets compensated by a single complex scalar at the
end of the $D$-term inflation, there are topologically stable D-strings.
These strings are topologically stable because of the non-trivial
homotopy $\pi_1$ of the vacuum of the broken $U(1)$ symmetry. This
manifold is $S_1$ and hence it contains  uncontractible  loops that can
be labeled by an integer $n$. This fact  guarantees the stability of the
cylindrically symmetric Higgs configurations when the phase of the Higgs
field winds by $2\pi n$ around some axis $\phi\, =\, \sqrt{\xi}\, {\rm
e}^{\rmi n\theta}$.

However, this is only true as long as $\phi$ is {\it not} transforming
under any other non-abelian symmetry group. A priori, there is nothing
that forbids such a transformation, and in fact in view of the
conjectured D-brane connection \cite{Dvali:2003zh}, such a situation is
very likely.  So let us see what will happen with the topological
stability of the $D$-term strings in the case where $\phi$ is  in a
representation of a larger symmetry group. For simplicity, let us assume
that $\phi$ transforms as a doublet of some gauge $SU(2)$ symmetry. This
symmetry combined together with our $D$-term $U(1)$ promotes the full
gauge group into $SU(2)\times U(1)$. When $\phi$ condenses, compensating
the $D$-term, the group is broken down to $U(1)'$, with the vacuum
manifold now being $SU(2)\times U(1)/U(1)$. This manifold is a three
sphere, and $\phi$ can take an expectation value at any point on it.
However,  unlike $S_1$, any closed loop on $S_3$ can be continuously
contracted into a point. Hence strings are no more stable, and unwind
without any topological obstruction.
 In such a scenario no cosmic strings would form in the phase transition after inflation.

Interestingly, to destabilize the $D$-term strings, the existence of a
non-abelian gauge group is strictly speaking  unnecessary. All we need is
that there is at least one more complex field $\phi'$ that carries
exactly the  same charge under the $D$-term $U(1)$ as $\phi$ does, and
that the value of this field is not fixed by the couplings in the
superpotential. In such a case, the global structure  of the vacuum
manifold is again $S_3$, although the gauge structure is $S_1$. That is,
there is an accidental {\it global} $U(2)$ symmetry. When the gauge group
breaks down to nothing by the vev of $\phi$, the global group breaks down
to $U(1)$. What matters for the topological stability is precisely the
global structure of the vacuum. The $D$-term string now can unwind
without any cost of potential energy. In other words, in this situation
the $D$-term strings become semi-local
strings~\cite{Vachaspati:1991dz}.\footnote{Before we submitted this
paper, \cite{Urrestilla:2004} appeared which also uses this point.} We
must stress however, that it may still cost a finite gradient energy to
unwind such a string, and hence they may still play some interesting role
in cosmology. This issue is beyond the scope of the present work.

\subsection{D-string instability as $D$-term string instability.}
\label{ss:DDinstab}

 We wish now to show that according to the conjecture of \cite{Dvali:2003zh},
the above discussed topological instability of the $D$-term strings can
be viewed as the corresponding instabilities of D-brane strings of string
theory \cite{Copeland:2003bj}. It is known that D$_1$ strings become
topologically unstable in the presence of D$_3$-branes.  That is, if the
D$_1$ is placed on top of a D$_3$ it can ``dissolve'' in the D$_3$ brane.
If the D$_1$ is placed at some distance apart, there is a tunneling
process by which it can break with two ends attached to the D$_3$ brane.
These attached ends look like monopoles on the D$_3$-brane. To connect
this instability with the one of the $D$-term strings,  it is convenient
to view it in the following way. Think of the D$_1-\mbox{D}_3$ system as
being formed from the system of two D$_3$ and one $\bar{\mbox{D}}_3$
branes, after one of the D$_3$-branes got annihilated by the
$\bar{\mbox{D}}_3$. This annihilation proceeds via tachyon condensation,
which cancels the energy of D$_3-\bar{\mbox{D}}_3$ pair.  This energy
breaks all the supersymmetries, and according to \cite{Dvali:2003zh} it
is in form of the $D$-term. Consider first a situation when all the
parent branes are on top of each other. The gauge symmetry of the
original system is then $U(2)\times U(1)$, and the tachyon transforms as
$(2,-1)$ under it.  This is identical to our example with $SU(2)\times
U(1)$  symmetry , in which the role of the $D$-term-compensating field is
assumed by the tachyon.\footnote{On a compact space, there should be
something else that absorbs RR and gravitational fluxes of the remaining
D$_3$-brane, in order to keep the tachyonic vacuum, without the $D$-term,
flat.  This does not affect our discussion.} The instability of the
string is clear in this language. The tachyonic vacuum is topologically
trivial, and so is the vacuum with the canceled $D$-term. So  $D$-term
strings are topologically unstable and can unwind, spreading flux out. In
the string language the same process is seen as dissolving D$_1$ string
on the D$_3$-brane, with its flux spreading out.

 Now let us displace one of the original D$_3$-branes in the perpendicular direction.
This Higgses the original symmetry to $U(1)\times U(1)\times U(1)$  and
there are monopoles in this system. In the language of four-dimensional
supergravity this is equivalent to giving a vev to the $SU(2)$-adjoint
Higgs.  After the tachyon condenses and compensates the $D$-term, the
symmetry is broken to $U(1)\times U(1)$, and strings can be formed.
However, these strings are not really stable but can break into monopole
anti-monopole pairs.

 We see that there is a complete correspondence with the $D$-term case.

\section{FI $D$-terms from D-branes, and moduli stabilization}
\label{ss:FIDfromD}

 The connection suggested in \cite{Dvali:2003zh} between the
$D$-term strings and BPS $D$-brane strings  ($D_{1+q}$-branes of type
$II$ string theory, wrapped on $q$-cycles) allowed  us to view many
properties of non-BPS brane-anti-brane systems from the point of view of
FI $D$-term supersymmetry breaking. Let us briefly review some of our
conclusions. Consider a pair of D$_{3+q} -\bar{\mbox{D}}_{3+q}$ branes
wrapped on a $q$-cycle of radius  $R$. We shall assume that $6-q$
remaining additional dimensions are also compactified on a cycle of
radius $R_{\bot}$. The D$_{3+q} -\bar{\mbox{D}}_{3+q}$ system breaks all
the supersymmetries. The low-energy gauge group consists of two
$U(1)$-symmetries. Their field strengths couple with opposite sign to the
Ramond-Ramond  $(2+ q)$-form $C_{(2+q)}$.  Hence, we can choose two
orthogonal combinations of these $U(1)$-s according to their RR
couplings. We shall be interested in the diagonal $U(1)$ (from now on we
shall call it simply $U(1)$) that has a non-zero RR charge. We shall
denote the corresponding two-form field strength by $F_{(2)}$. Since the
net RR charge of D$_{3+q} -\bar{\mbox{D}}_{3+q}$ is zero, the system is
unstable towards annihilation. Annihilation can be described as
condensation of the open string tachyon $\phi$, which Higgses the
diagonal $U(1)$-symmetry. The tachyonic vacuum is a closed string vacuum
with no D$_{3+q}$ branes.

 In our $D$-term description, the above stringy picture translates as follows.
The supersymmetry breaking by the D$_{3+q} -\bar{\mbox{D}}_{3+q}$-system
is seen in effective $4d$, $N=1$ supergravity as breaking by the FI
$D$-term. In the limit $R_{\bot} \rightarrow \infty$, the world-volume
$4d$ supergravity approaches the rigid limit, and the FI term is related
to the brane tension and the radius of a $q$-cycle in the following way
\begin{equation}
\label{tensions1} 2(2\pi R)^q T_{3+q} \, = \, {2 R^q \over g_s(2\pi)^3
\alpha^\prime{}^{q+4\over 2}} \, = \, {g^2 \over 2} \xi^2\,,
\end{equation}
where $g_s$ and $g$ are the string and the world-volume gauge couplings,
respectively.

Notice that the above form is valid in the limit in which gravity is
decoupled ($M_P = \infty$). In this limit, the FI term cannot drive any
$4d$ $D$-term inflation. After taking the finite compactification volume
and bringing the $4d$ action to the Einstein form by Weyl rescaling, the
effective four-dimensional source will, of course, become dependent on
the volume modulus, in accordance to \cite{Burgess:2003ic}, and in order
to get inflation one has to address the issue of the volume
stabilization. However, at the moment we shall work in the
infinite-volume limit, in which the low-energy world-volume theory
reduces to a gauge theory with the above FI term.

After annihilation of branes, this $D$-term is compensated by the vev of
the tachyon $\phi$, which corresponds to a scalar component of a chiral
superfield. The tachyonic vacuum is a supersymmetric vacuum with
vanishing $D$-term. When branes are far apart, the mass$^2$ of the
tachyon becomes positive, and the $D$-term is non-zero. As suggested in
\cite{Dvali:2003zh}, the tachyon superpotential is
\begin{equation}
\label{Wtachyon} W \, = \, \phi_0 \phi\bar\phi\,,
\end{equation}
where $\bar{\phi}$ is the supersymmetric tachyonic partner, a chiral
superfield with opposite $U(1)$-charge. $\phi_0$ is a $U(1)$-neutral
superfield that corresponds to the inter-brane separation ($r$)
\begin{equation}
\label{phir} \phi_0 \, = \, M_s^2 r \,,
\end{equation}
where $M_s$ is the string scale. Although there are other fields with
masses comparable to tachyon, we have only focused our attention to the
superfields whose chiral fermionic components become massless at zero
brane separation. These are fermionic partners of $\phi_0,\, \Phi,\,
\bar{\phi}$ as well as gaugino and $4d$ gravitino. The latter is
decoupled in the infinite compactification volume limit. There are also
massless fermionic modes corresponding to the center of mass motion,
which are not of our interest. All other fermions are massive. Note also
that fermionic partners of $\phi,\,\bar{\phi}$ are massless only at the
zero separation point ($\phi_0 = 0$).

Hence, we see that in the rigid limit, the low-energy dynamics of
(D$_{3+q} -\bar{\mbox{D}}_{3+q}$)-system is described by the globally
supersymmetric limit of the $D$-term inflation model discussed in
section~\ref{s:DtermcstFI}. We now wish to consider the case of finite
$M_P$ (that is finite $R_{\bot}$). A useful guideline for understanding
what may happen in the finite-volume case is the cancellation of the
$U(1)$-gauge anomaly.
First let us observe that in the rigid limit,  there is no
$U(1)$-anomaly, since the gaugino  and the fermionic partner of $\phi_0$
are neutral, the $4d$-gravitino is decoupled ($M_P=\infty$) and the
anomalies of the chiral fermionic partners of $\phi$ and $\bar{\phi}$
exactly cancel. Now, for finite $M_P$, the situation may change and the
charges of the chiral fermions may shift. One supergravity example of
such a situation when going to finite $M_P$ re-arranges charges of the
chiral fermions was discussed in section~\ref{s:DtermcstFI}. In that
example, the charges of chiral fermions shift as described in
(\ref{tildeq}),  and the gaugino and gravitino acquire charges equal to
$G$.  However, with shifted fermionic charges the chiral gauge anomaly is
non-vanishing anymore, and is equal to the gravitino contribution
according to (\ref{C}). Due to this fact, the analogous charge-shift in
the present case may create a seeming puzzle. Starting from anomaly-free
high-dimensional theory, we cannot create an anomaly by simply changing
the compactification radius! So consistency of the compactification
requires that if there is such an anomalous shift in fermionic charges,
there must be an axion whose shift exactly cancels the anomaly by the GS
mechanism. The axion that is charged under $U(1)$ indeed exist in the
theory in form of the RR axion, as we shall now demonstrate.

 The $U(1)$ field strength has the following world-volume coupling to the RR
$(2+q)$-form
\begin{equation}
g_s(2\pi\alpha')T_{(3+q)} \int_{3 + 1 + q} F_{(2)}\wedge C_{(2 + q)}\,.
\label{rrcoupling}
\end{equation}
The $4d$ dual of the $4d$-zero mode of $C_{(2+q)}$ is precisely the axion
that shifts under $U(1)$.  To see this, let us concentrate on the
components of $C_{(2+q)}$ that take only two indices in the four
non-compact dimensions, and the rest on a $q$-cycle. These components
effectively define a two form, which we can call $C_{(2)}$. We shall only
be interested in a $4d$ zero mode of this two-form (higher KK-modes are
irrelevant for the anomaly cancellation). We can now go into the dual
description of the $C_{(2)}$-form in terms of an axion
\begin{equation}
\label{cduala} \rmd C_{(2)} \rightarrow *\,\rmd a\,,
\end{equation}
where star denotes a $4d$ Hodge-dual. Under this duality transformation
we have to replace
\begin{equation}
\label{c-a} (\rmd C_{(2)})^2  \rightarrow  (\rmd a \, - \,g Q_a W)^2\,,
\end{equation}
Under the $U(1)$, the axion shifts as
\begin{equation}
\qquad a \rightarrow a \, + \,g Q_a \alpha(x)\,.
 \label{deltaa}
\end{equation}
$Q_a$ can be found from (\ref{rrcoupling}) by integrating over the extra
coordinates and using the relation (\ref{tensions1}) between $\xi$ and
the brane tension, as well as the relation  $g^2 = 8\pi g_s
{\alpha^\prime{}^{q / 2}/ R^q}$.  The result\footnote{As shown in
\cite{Dvali:2003zh}, this is also the right value that correctly
reproduces the long range RR field of the $D$-term string formed by the
tachyon.} is $Q_a = \xi/M_P^2$.  Hence, we see that in the $4d$ theory
there is always the axion with right transformation properties, which
could potentially cancel the anomaly appearing from the chiral fermion
sector, provided there is a coupling
\begin{equation}
 a \, F \wedge F\,.
 \label{aFF}
\end{equation}
Existence of such a coupling is in general compactification
dependent\footnote{We thank Juan Maldacena for discussions on this and
other issues.} and hence so must be the resulting fermionic charge
assignment. In case that it is absent, the anomaly of the chiral fermions
must be separately zero. When such coupling is present, the chiral
fermion set must be anomalous, and exactly compensate the anomaly
resulting from the shift of the axion via the GS mechanism. Hence, in
such case the effective $4d$ description of D$-\bar{\mbox{D}}$ systems
will be in terms of the ``anomalous'' $U(1)$ symmetry.

 Let us finally tress that the existence of the axion that shifts under
the $U(1)$ gauge symmetry has important consequences for the issue of
moduli-stabilization.

In the $4d$ theory, the axion becomes a lowest component of a chiral
supermultiplet $\Phi  \, = \sigma + \rmi a$, where the role of $\sigma$
is model dependent. For instance, in type $II$ B theories, $\sigma$ can
either be some combination of the dilaton and the volume modulus, or an
NS-NS two form. The fact that the axion shifts under $U(1)$, implies that
(in $4d$ Einstein frame) the FI term is $\Phi $-dependent. This of course
raises the issue of the stabilization of the real part of $\Phi $.
Stabilization could in principle happen via some non-perturbatively
generated superpotential. However, because of the $U(1)$-symmetry, the
superpotential must transform homogeneously under the shift of $\Phi$,
that is, be invariant up to the K{\"a}hler transformation. Hence, up to an
arbitrary K{\"a}hler transformation the $\Phi$-dependence of the
superpotential can only be through the invariant(s) of the form
\begin{equation}
{\rm e}^{{c\over Q_a}\Phi} w_0\,,
\end{equation}
where $c$ is a constant and $w_0$ is some holomorphic function of the
chiral superfields (in our case $\phi_0, \phi, \bar{\phi}$) that carries
the overall charge $Q=-c$ with respect to $U(1)$. We see that holomorphy
and $U(1)$ invariance can potentially severely constrain the form of the
stabilizing potential for $\Phi$.

\section{Conclusion}

In this paper we have clarified the status of constant FI terms $\xi$ in
$N=1$, $d=4$ supergravity in general and in examples. Their presence
shows up in covariant derivatives of all fermions and in the
supersymmetry transformation laws, since the relevant local $U(1)$
symmetry is a gauged $R$-symmetry. These new couplings proportional to
${g\xi/ M_P^2}$  lead to gauge and gravitational anomalies. Under certain
conditions it is possible to cancel the gauge anomalies.

One of the important restrictions on supergravity with constant FI terms
is the following: the superpotential $W$  has to  transform under $U(1)$
gauge symmetry, $\delta W= -\rmi{g\xi\over M^2_{P}} W$, otherwise the
constant FI term $\xi$ has to vanish. This requirement is consistent with
the fact that in the gauge theory at $M_{P}\rightarrow \infty$ the
potential is $U(1)$ invariant. However, we consider local supersymmetry
of the classical supergravity action, in which terms of the order ${g\xi/
M_P^2}$ are all taken into account.


In the  example of a $D$-term inflation \cite{Binetruy:1996xj}, it is
possible to generalize the original model with rigid supersymmetry to
exact local supersymmetry. A gauge theory potential of the $D$-term model
$W=\phi_0 \phi_+\phi_-$ is neutral under $U(1)$ symmetry in  gauge theory
with constant FI terms. In this paper we  have found how to promote  this
model to the supergravity level with constant FI terms: we required that
the total charge of $\phi_+$ and $\phi_-$ fields does not vanish but is
equal to $-{\xi/ M_P^2}$. The gauge theory anomaly from gravitino,
gaugino, original chiral multiplets $\phi_0$, $\phi_+$ and $\phi_-$ and
additional 3 chiral multiplets  can be cancelled.

In string theory there are no known examples of constant FI terms $\xi$.
Only moduli-dependent $D$-terms are available
\cite{Dine:1987xk,Burgess:2003ic}. In absence of constant FI terms, the
rules of local supersymmetry require the superpotential to be invariant
under the $U(1)$-gauged symmetry (for invariant K{\"a}hler potential). In
such theories, where the cancellation of gauge anomaly is due to the
Green-Schwarz mechanism with the shift of the axion and the coupling
$aFF^*$, one has to stabilize the scalar partner of the axion to get the
effective supergravity with constant FI terms. Some efforts in this
direction have been made recently in
\cite{Kachru:2003aw,Kachru:2003sx,Burgess:2003ic,Hsu:2003cy,Firouzjahi:2003zy,Angelantonj:2003zx,Koyama:2003yc}.
It is possible that a stringy version of the D-brane inflation
\cite{Dvali:1998pa} with improvement with respect to volume and dilaton
stabilization will be derived in the future and that the problems with
inflation in string theory, pointed out in \cite{Kachru:2003sx}, will be
resolved. This kind of string cosmology program requires a better
understanding of the structure of 3+1 dimensional $N=1$ supergravity with
constant FI terms as well as the one with field-dependent $D$-terms. This
paper has clarified the properties of such theories.

In this paper we have investigated another interesting role that FI terms
can play in string theory. This role is based on the recently-suggested
\cite{Dvali:2003zh} equivalence between the $4d$ supergravity $D$-term
strings and D-branes of type $II$ string theory. According to it,
brane-anti-brane systems in an effective $4d$ theory can be viewed as
gauge theories with non-zero FI term, in which the axion shifting under
the $U(1)$-symmetry comes from the RR sector. The tachyonic instability
of the brane-anti-brane system is seen as the instability triggered by
the FI-term. Thus, many important properties of D$-\bar{\mbox{D}}$
systems can be understood in the light of $4d$ supergravity with FI
$D$-terms. Certain aspects of stability of some string compactifications
with branes and anti-branes can be understood in the language of
supergravity vacua with non-zero $D$-terms. For instance, the shift of
the axion under the gauge $U(1)$ symmetry gives selection rules for the
possible invariants of the stabilizing superpotential.

 On the cosmic string front, we have provided some additional
consistency checks of the conjectured correspondence \cite{Dvali:2003zh}
between $D$-term-strings and D-branes, by mapping the instabilities of
the two.

 Finally, we have studied in non-minimal model
the zero mode content on the BPS $D$-term cosmic string solutions of
$N=1$ supergravity with constant FI terms \cite{Dvali:2003zh}. We have
discovered a puzzling property that the numbers of bosonic and fermionic
zero modes can be {\it arbitrarily} different. We have explained this
puzzling behaviour by unusual properties of unbroken supersymmetry.

\subsection*{Acknowledgments}
It is a pleasure to thank N. Arkani-Hamed, M. Douglas, D. Freedman, G.
Gibbons, S. Kachru, A. Linde, D. Lyth, J. Maldacena, R. Myers, K. Stelle,
P. West and B. de Wit for useful discussions.  The work of R.K. was
supported by NSF grant PHY-0244728. The research of G.D. is supported in
part by a David and Lucile Packard Foundation Fellowship for Science and
Engineering, and by the NSF grant PHY-0070787. The work of A.V.P. was
partially supported by the European Community's Human Potential Programme
under contract HPRN-CT-2000-00131 Quantum Spacetime and  in part by the
Federal Office for Scientific, Technical and Cultural Affairs, Belgium,
through the Inter-university Attraction Pole P5/27. R. K. and P. B.  are
grateful to the organizers of Santa Barbara string cosmology workshop,
and R.K., P.B. and A.V.P. to the organizers of the workshop in Bad Honnef
on dark matter and dark energy, where part of this work was performed. P.
B., G. D. and A.V.P. thank the Institute of theoretical physics of
Stanford for the hospitality.

\newpage
\appendix

\section{Residual superalgebra of the $D$-term string}
\label{app:decompsusy}

The commutator between two supersymmetry transformations is given by
\begin{equation}
  \left[ \delta (\epsilon _1),\delta (\epsilon _2)\right] =\ft12 \delta
  _{\rm CGCT}\left(\xi ^\mu (\epsilon _1,\epsilon _2)\right)\,,\qquad
\xi ^\mu (\epsilon _1,\epsilon _2)\equiv \bar \epsilon _2 \gamma ^\mu
\epsilon _1\,,
 \label{algebraeps}
\end{equation}
where the covariant general coordinate transformation (CGCT) is a
combination of general coordinate transformations, Lorentz
transformations and gauge transformations (at least the bosonic part):
\begin{equation}
  \delta_{\rm CGCT}(\xi ^\mu )= \delta_{\rm GCT}(\xi ^\mu )- \delta _{\rm
  Lor}(\xi ^\mu \omega _\mu ^{ab}) -\delta _{\rm G}(\xi ^\mu W_\mu )\,.
 \label{CGCT}
\end{equation}
We have e.g. on the scalars and the vectors
\begin{equation}
  \delta_{\rm CGCT}(\xi ^\mu )\phi =\xi ^\mu \hat{\partial  }_\mu \phi\,,\qquad
   \delta_{\rm CGCT}(\xi ^\mu )W_\mu= \xi ^\nu F_{ \nu\mu }\,.
 \label{exCGCT}
\end{equation}
Introducing notations
\begin{equation}
  \delta (\epsilon )= \bar \epsilon Q= \epsilon ^\alpha Q_\alpha \,,\qquad
\delta_{\rm GCT}(\xi ^\mu )=\xi ^\mu P_\mu\,,\qquad \delta _{\rm
  Lor}(\Lambda  ^{ab})= \ft12\Lambda ^{ab}L_{ab}\,, \qquad \delta _{\rm G}(\alpha
  )=\alpha T\,,
 \label{defgenerators}
\end{equation}
we can write the supersymmetry algebra as
\begin{equation}
  \left\{ Q_\alpha ,Q_\beta \right\} =\ft12(\gamma ^\mu {\cal C}^{-1})_{\alpha \beta }
  \left( P_\mu -\omega _\mu ^{12}L_{12}-W_\mu T\right) .
 \label{algebraQ}
\end{equation}

The split between the 4 initial supersymmetries that is relevant for the
$D$-term strings occurs through the projectors
\begin{equation}
  \Pi _\pm = \ft12\left( \unity \pm \rmi \gamma _5 \gamma ^{12}\right).
 \label{Pipm}
\end{equation}
These projectors satisfy
\begin{equation}
  \gamma ^1 \Pi _\pm =\Pi _{\pm }\gamma ^1\,,\qquad\gamma ^2 \Pi _\pm =\Pi _{\pm }\gamma
  ^2\,,
 \label{commPigamma}
\end{equation}
and Majorana conjugates exchange the projections, e.g.
\begin{equation}
  \overline{\Pi _\pm \epsilon}= \overline{\epsilon }\Pi _\mp\,.
 \label{barpm}
\end{equation}
This implies that $\xi ^\mu (\epsilon _1,\epsilon _2)$, for $\mu=1,2$, is
only nonzero for two spinors of different $\pm $-type. Using notations as
$\epsilon _\pm=\Pi _\pm\epsilon $, we have
\begin{equation}
  \xi ^\mu (\epsilon _1,\epsilon _2)=\bar \epsilon _{2+}\gamma^\mu \epsilon
  _{1-}+\bar \epsilon _{2-}\gamma^\mu \epsilon
  _{1+}\qquad\mbox{for }\mu =1,2\,.
 \label{ximu+-}
\end{equation}

In terms of supersymmetry generators, this implies that the only nonzero
anticommutators are
\begin{equation}
   \left\{ Q_{+\alpha} ,Q_{-\beta} \right\} =\ft12(\Pi _+\gamma ^\mu {\cal C}^{-1})_{\alpha \beta }
  \left( P_\mu -\omega _\mu ^{12}L_{12}-W_\mu T\right).
 \label{Q+Q-}
\end{equation}

\section{Relating bosonic and fermionic modes} \label{app:BosFerModes}

In this appendix, we will repeat  the argument why in supersymmetric
theories one can construct a fermionic mode for every bosonic mode and
vice versa. This leads to the theorem about the equality of the number of
fermions and bosons in a supersymmetric theory. This expos{\'e} will clarify
that the theorem depends crucially on the algebra, as the fact that the
fermion mode constructed out of a bosonic mode and vice versa could be
degenerate if the algebra is not suitable.

The construction starts from an expansion around a solution of the theory
where the bosonic fields $\phi ^i$ have a value $\phi_0^i$ and the
fermions $\psi ^a$ are zero $\psi _0^a=0$. It will use the concept of
residual supersymmetry as exposed in~\cite{Kallosh:1998ug}, and the
notations are taken from the final pages of that paper. We consider a
supersymmetry transformation of the bosons and fermions, which, using the
condensed notation of B. DeWitt~\cite{DeWitt:1967ub}, can be written as
\begin{equation}
  \delta (\epsilon )\phi ^i = R^i{}_\alpha(\phi ,\psi ) \epsilon
  ^\alpha\,,\qquad\delta (\epsilon )\psi ^a = R^a{}_\alpha(\phi ,\psi ) \epsilon
  ^\alpha\,.
 \label{susycondensed}
\end{equation}
The index $\alpha $ represents only the residual supersymmetry at the
solution of the theory, which implies $R^a{}_\alpha(\phi_0 ,0 )=0$. We
also have $R^i{}_\alpha(\phi_0,0 )=0$, which is already obvious from the
fact that this is a fermionic quantity and thus vanishes for zero
fermions. We write the fields as
\begin{equation}
  \phi ^i=\phi ^i_0+ \Delta \phi ^i\,, \qquad \psi ^a= \Delta \psi ^a\,,
 \label{Deltafields}
\end{equation}
and expand the action up to second order in the perturbations $\Delta
\phi ^i$ and $\Delta \psi ^a$:
\begin{eqnarray}
  &&S= S^0 + \ft12 \Delta \phi ^iS_{ij}^0\Delta \phi ^j +\ft12 \Delta \psi ^aS_{ab}^0\Delta \psi
  ^b+ {\cal O}(\Delta ^3)\,, \nonumber\\
  &&S^0\equiv S(\phi ^i_0,0)\,, \qquad S_{ij}^0\equiv  \frac{\delta }{\delta \phi ^i} \frac{\delta }{\delta \phi ^i}
  S(\phi ^i_0,0)\,, \qquad S_{ab}^0\equiv \frac{\stackrel{\rightarrow} {\delta}}{\delta \psi ^a}
   \frac{\stackrel{\leftarrow}{\delta} }{\delta \psi ^b}
  S(\phi ^i_0,0)\,.
 \label{Sexpanded}
\end{eqnarray}
First order perturbations are absent because the zero state is a solution
of the classical field equations, and mixed terms are absent due to their
fermionic nature and $\psi ^a_0=0$. As the supersymmetry transformations
on the vacuum states are zero, supersymmetry acts only on the
perturbations and in first order is
\begin{equation}
  \delta (\epsilon )\Delta \phi ^i=R^i{}_{\alpha,a}(\phi_0,0 )\Delta \psi ^a\epsilon
  ^\alpha + {\cal O}(\Delta ^2)\,, \qquad
\delta (\epsilon )\Delta \psi ^a=R^a{}_{\alpha,i}(\phi_0,0 )\Delta
\phi^i\epsilon^\alpha + {\cal O}(\Delta ^2)\,,
 \label{susyDelta}
\end{equation}
where the notations $,i$ and $,a$ denote derivatives of the
transformation laws. Again, other terms are absent due to $\psi ^a_0=0$.
The statement that the action is supersymmetric is
\begin{equation}
 0= \delta  (\epsilon )S=\left[ \Delta \phi ^iS_{ij}^0R^j{}_{\alpha,a}(\phi_0,0 )\Delta \psi ^a
   +  \Delta \psi ^aS_{ab}^0R^b{}_{\alpha,i}(\phi_0,0 )\Delta
\phi^i\right] \epsilon^\alpha + {\cal O}(\Delta ^3)\,.
 \label{delSDelta}
\end{equation}
Consider now a solution of the fermionic field equations $ \Delta \psi
^aS_{ab}^0=0$ and consider arbitrary $\Delta \phi ^i$. Then we find
\begin{equation}
  S_{ij}^0R^j{}_{\alpha,a}(\phi_0,0 )\Delta \psi ^a =0\,,
 \label{BosfromFerm}
\end{equation}
which shows that
\begin{equation}
  \Delta \psi^a \mbox{ solution of field eqs.}\quad \rightarrow \quad\phi^i (\epsilon
  )=R^i{}_{\alpha,a}(\phi_0,0 )\Delta \psi ^a\epsilon ^\alpha\mbox{ solution of field eqs.}
 \label{fermtoboson}
\end{equation}
for any $\epsilon $ that parametrizes a residual supersymmetry.
Similarly, for any solution of the bosonic field equations one constructs
a solution of the fermionic field equations as
\begin{equation}
  \Delta \phi ^i \mbox{ solution of field eqs.}\quad \rightarrow \quad
\psi ^a(\epsilon ) = R^a{}_{\alpha,i}(\phi_0,0 )\Delta \phi ^i\epsilon
^\alpha\mbox{ solution of field eqs.}
 \label{bostofermion}
\end{equation}
For counting the number of bosonic and fermionic modes, one may imagine
that these mappings can be either degenerate or not. However, applying
the map twice on a bosonic mode $\Delta \phi ^i$ gives that there should
be a bosonic solution of the field equations
\begin{equation}
  \tilde \Delta \phi ^i(\epsilon _1,\epsilon _2)=
  R^i{}_{\alpha,a}(\phi_0,0 )R^a{}_{\beta,j}(\phi_0,0 )\Delta \phi ^j\epsilon _2^\beta \epsilon _1^\alpha \,.
 \label{solnBosfromBos}
\end{equation}
The involved square of the transformation operators is what appears at
first order in the anticommutator of the supersymmetry with parameter
$\epsilon _1$ with the one with parameter $\epsilon _2$. Therefore,
\textit{if this anticommutator is non-degenerate}, then this gives an
invertible mapping between bosonic states. This is only possible if the
intermediate states are non-degenerate and thus there are at least as
many fermionic modes as bosonic modes. Starting the argument with the
fermions, we arrive at the statement that there are at least as many
bosonic modes as fermionic modes. This gives the theorem that the number
of bosonic modes and fermionic modes are equal.

This counting depends on the invertibility of the square of
supersymmetry. The latter gives in the simple cases just the energy (or
time derivative of the fields) such that the theorem applies. In some
cases the algebra involves gauge transformations, and the theorem only
applies to the gauge-invariant states. In our case the supersymmetries
are nilpotent, and the theorem does not apply. In fact the
construction~(\ref{zeroshift}) can be seen to give a vanishing bosonic
mode.

\providecommand{\href}[2]{#2}\begingroup\raggedright\endgroup


\end{document}